\documentclass[11pt]{article}
\usepackage{jheppub}

\usepackage{amsmath}
\usepackage{amssymb}
\usepackage{verbatim}

\usepackage{empheq}

\usepackage{dsfont}

\usepackage[normalem]{ulem}

\usepackage[T1,OT1]{fontenc}

\usepackage{graphicx}

\allowdisplaybreaks[4]

\usepackage[hang]{footmisc}
\renewcommand{\d}{\mathrm{d}}
\newcommand{\e}{\mathrm{e}}
\newcommand{\w}{\wedge}

\newcommand{\nl}{\notag \\ &\quad\,}
\newcommand{\nll}{\notag \\ &}

\title{$D$-term and $F$-term potential of non-Abelian D7-branes in IIB Calabi-Yau orientifolds}

\author{Daniel Junghans}

\affiliation{Max-Planck-Institut f\"ur Physik (Werner-Heisenberg-Institut), Boltzmannstr.~8, 85748 Garching,
Germany}

\emailAdd{junghans @ mpp.mpg.de}

\notoc

\abstract{We derive the 4d $D$-term and $F$-term potential generated by a D7-brane stack
in Calabi-Yau orientifold compactifications of type IIB string theory, taking into account the effect of worldvolume and ISD bulk fluxes and the non-Abelian nature of the brane fields.
This potential is an important ingredient for moduli stabilization in IIB, e.g., in de Sitter scenarios using a T-brane uplift. While aspects of the $D$-term and $F$-term parts of the potential have been inferred before with various arguments, we provide here a unified first-principles derivation of the full potential by dimensionally reducing the non-Abelian DBI action of the D7-branes. Our result generalizes previous expressions and  agrees with them where applicable, thus confirming the mutual consistency of the various approaches.
Moreover, we provide a thorough discussion of corrections from integrating out KK modes and argue that demanding control over such corrections implies bounds on the Calabi-Yau volume and the brane fields.
Aside from these various results for the 4d effective action, our calculation also shows how bulk fluxes correct the potential and the BPS equations in the 8d worldvolume theory that governs the microscopic dynamics of non-Abelian branes.

}

\begin{document}

\begin{flushright}
MPP-2026-182
\end{flushright}

\numberwithin{equation}{section}

\maketitle

\newpage

\section{Introduction}

A long-standing problem in string theory is to determine which 4d low-energy effective field theories can arise from its compactifications.
This is particularly challenging for compactifications preserving little or no supersymmetry since they are not protected against perturbative and non-perturbative corrections. On the other hand, such compactifications are precisely the ones which have a chance of admitting (semi-)realistic vacua with full moduli stabilization and are thus the most interesting from the point of view of phenomenology.
A much studied
class of models in this context is given by Calabi-Yau orientifolds in type IIB string theory with O3/O7-planes, D3/D7-branes and 3-form fluxes, which are described by 4d $\mathcal{N}=1$ supergravities at low energies \cite{Giddings:2001yu, Grimm:2004uq}. How moduli stabilization works in this setting and which types of vacua can arise has been intensely debated for over two decades, sparked by the famous scenarios proposed in \cite{Kachru:2003aw, Balasubramanian:2005zx}. See, e.g., \cite{Denef:2008wq, McAllister:2023vgy} for reviews.

In order to address such questions,
the crucial quantity one has to compute is the scalar potential,
which generically receives contributions from both $D$-terms and $F$-terms.
The present paper is concerned with a specific part of this potential, namely the one generated by a stack of D7-branes wrapped on a divisor $D$.
We derive the full $D$-term and $F$-term potential of the branes (at the order $\alpha'^2$ and in the probe regime), for a general Calabi-Yau orientifold compactification including worldvolume and ISD bulk fluxes and taking into account that the brane moduli can be non-Abelian.

The main motivation to study the non-Abelian potential is that, in the presence of worldvolume flux, D7-branes can form non-commutative bound states known as T-branes \cite{Donagi:2003hh, Cecotti:2010bp}. Such configurations have various interesting consequences as discussed, e.g., in \cite{Cecotti:2010bp}.
In the context of moduli stabilization, an intriguing proposal made in \cite{Cicoli:2015ylx} is that T-branes can provide an uplift to de Sitter vacua in the framework of the LARGE-volume scenario \cite{Balasubramanian:2005zx}. Such an uplift mechanism, if reliable, could be a promising alternative to the anti-brane uplift whose validity is by now in serious doubt (since there is strong evidence that either warping or curvature corrections lead to a loss of control there \cite{Carta:2019rhx, Gao:2020xqh, Carta:2021lqg, Junghans:2022exo, Gao:2022fdi, Junghans:2022kxg, Blumenhagen:2022dbo, Hebecker:2022zme, Schreyer:2022len, Schreyer:2024pml, McAllister:2024lnt, ValeixoBento:2023nbv}). We will provide a detailed analysis of the T-brane uplift in our upcoming paper \cite{Junghans:2026hfm}.
The goal of the present paper is to perform a careful derivation of the brane potential, which is a crucial prerequisite of the T-brane-uplifting scenario -- and, more generally, for any attempt to embed T-branes into a globally consistent compactification with stabilized moduli.

Many works have already studied aspects of the D7-brane potential before, both in the Abelian and non-Abelian case.
The $D$-term part was obtained for a single brane in \cite{Jockers:2004yj, Jockers:2005zy, Haack:2006cy, Martucci:2006ij} and partial results for a non-Abelian brane stack were found in \cite{Marchesano:2010bs, Marchesano:2017kke, Cicoli:2015ylx, Marchesano:2016cqg} using various different approaches.
For example, the method employed in \cite{Jockers:2005zy} is based on a dimensional reduction of the fermionic terms in the DBI action, while \cite{Haack:2006cy} instead reduces the bosonic DBI action. In \cite{Martucci:2006ij}, the $D$-terms are inferred using the 8d supersymmetry equations in the pure-spinor formalism and an argument based on the action of cosmic strings. The calculation in \cite[App.~A]{Marchesano:2017kke} takes as a starting point the 8d BPS equations of \cite{Beasley:2008dc}, which are motivated by an argument that the 8d worldvolume theory should be a twisted Super-Yang-Mills theory. The approach in \cite{Marchesano:2010bs} is instead based on a suggestive analogy with the Chern-Simons action observed in \cite{Butti:2007aq}. Yet another way is to exploit that the structure of 4d $\mathcal{N}=1$ supergravity fixes the $D$-term potential in terms of the K\"ahler potential, the U$(1)$ charges of the various moduli and the gauge-kinetic function, see, e.g., \cite{Jockers:2004yj, Jockers:2005zy, Cicoli:2015ylx, Marchesano:2016cqg} for this approach.

Also the $F$-term part of the brane potential has been determined in the literature in multiple ways. In the Abelian case, the corresponding superpotential was found in \cite{Jockers:2005zy} by dimensionally reducing the fermionic DBI action and in \cite{Martucci:2006ij} using the pure-spinor formalism and arguments based on BPS domain walls. The most general expression was obtained in \cite{Denef:2008wq, Arends:2014qca} using F-theory, including a previously unnoticed backreaction term (see also \cite{Lust:2005bd}). 
Moreover, educated guesses for a non-Abelian generalization were made in \cite{Marchesano:2009rz, Marchesano:2010bs}.\footnote{There are also non-perturbative corrections to the brane superpotential \cite{Blumenhagen:2007sm, Marchesano:2009rz} but they will not be studied in this paper.}

While there is nothing wrong with these various approaches, many of them are somewhat indirect. Moreover, only partial results are available in the non-Abelian case, e.g., for the $D$-term potential. By consistency of string theory, the brane potential must be obtainable directly from a dimensional reduction of the non-Abelian DBI action constructed in \cite{Myers:1999ps}.
Such a calculation has so far only been carried out in the special case where the branes wrap a $T^4$ or K3 and assuming vanishing worldvolume flux \cite{Camara:2004jj}.
For an Abelian brane, a dimensional reduction was furthermore done in \cite{Haack:2006cy}, under the assumption that the deformation moduli and $(2,0)$-form worldvolume fluxes are set to zero.
In this paper, we will do the dimensional reduction in full generality, i.e., for divisors on a general Calabi-Yau orientifold, taking into account the non-Abelian nature of the branes and allowing non-vanishing brane moduli, worldvolume fluxes and bulk fluxes.

Our approach allows us to compute the $D$-term and $F$-term potential of the D7-branes \emph{in a single calculation}, starting from first principles.
The advantage of this is two-fold. First, it provides a generalization and independent verification of previous arguments and proposals.
Second, all numerical and bulk-moduli-dependent prefactors in the potential can be reliably computed at once using the same conventions. Previous results on the different parts of the potential are scattered over many different papers with different conventions and assumptions so that it is difficult to keep track of such factors. However, the precise values of the latter are essential to assess moduli-stabilization scenarios without parametric control such as the previously mentioned T-brane uplifting, as will be discussed in more detail in \cite{Junghans:2026hfm}.

It turns out that the calculation we perform
involves a number of subtleties that either do not occur in the Abelian case or are at least much easier to handle there.
First of all, the brane moduli do no longer commute so that one has to keep track of various commutators throughout the computation.
Another complication is that it is in general not possible to choose a K\"ahler metric which is block-diagonal between components along the divisor $D$ and transverse components that are constant on $D$. This makes it rather tedious to do a normal-coordinate expansion of the DBI action and extract the dependence on the deformation moduli.
Finally, to obtain the 4d effective action, one has to carefully integrate out various KK modes and estimate under which conditions the resulting corrections are under control.
In non-commutative backgrounds, this is complicated by the fact that worldvolume fluxes and commutators of brane fields are typically not harmonic \cite{Marchesano:2017kke}.

After the dust settles, our calculation yields the $D$-term and $F$-term potential at the order $\alpha'^2$ (i.e., including at most four brane fields or derivatives)
and at leading order in $g_s$ (i.e., in the probe regime ignoring backreaction corrections). The result we obtain is, to our knowledge, the most general available.
It furthermore reproduces previous results where applicable.\footnote{Aside from differences in numerical factors in the various papers, which may be due to different conventions.} Although one may object that the latter was expected, let us stress again that it provides a highly non-trivial consistency check. Indeed, it is quite satisfying to see that, after a number of non-trivial manipulations, our final expression for the potential greatly simplifies and agrees with those obtained in completely different approaches, e.g., based on the Chern-Simons action, F-theory or domain walls. To give one example, rewriting the DBI action involves a non-trivial cancelation between three terms at the order $\alpha'^2$ involving the curvature of the normal bundle
and two brane fields. Such $\sim \mathcal{R}\phi^2$ terms could in principle have contributed to the brane potential at the order $\alpha'^2$ but precisely cancel, as required by the consistency with previous results.

Apart from determining the brane potential in 4d, our calculation also sheds light on the \emph{microscopic} brane dynamics, i.e., the 8d parent theory on the worldvolume. In particular, we obtain how the 8d potential and the 8d BPS equations
are corrected in the presence of bulk fluxes. A correction to these equations is not surprising since, from the point of view of the 4d low-energy theory, it is well-known that bulk fluxes contribute to the superpotential \cite{Gukov:1999ya} and thus enter the 4d $F$-term equations. Hence, there should be a corresponding effect also in the 8d equations, which is indeed what we find. An interesting future application of these results would be, e.g., to study directly in 8d how T-brane solutions are affected by supersymmetry breaking in the bulk, including constraints on higher KK modes which are not visible in 4d.

This paper is organized as follows. In Section \ref{dbi}, we review the non-Abelian DBI action and state our conventions. In Section \ref{normal}, we perform a normal-coordinate expansion to determine the brane-moduli dependence of the action up to the order $\alpha'^2$. In Section \ref{exp}, we rewrite the action in a way convenient for our purpose and furthermore state the 8d potential and the 8d BPS equations.
In Section \ref{harmo}, we expand the action in harmonics and discuss corrections from integrating out KK modes. In Section \ref{4dd}, we derive the 4d $D$-term and $F$-term potential of the D7-branes as well as the K\"ahler potential and the superpotential. In Section \ref{concl}, we summarize and conclude.

\section{Non-Abelian DBI action}
\label{dbi}

Our starting point is the DBI action of a stack of $N$ D7-branes, which is given by \cite{Myers:1999ps}
\begin{equation}
S_\text{DBI}= - 2\pi g_s\, \text{STr} \int \d^8 x\, \sqrt{-\text{det}\left( (\iota^*_\phi E)_{\alpha\beta} + (\iota^*_\phi H)_{\alpha\beta} + \frac{F_{\alpha\beta}}{2\pi \sqrt{g_s}} \right)\, \text{det}(Q^m{}_n)} \label{myersdbi}
\end{equation}
in the Einstein frame with
\begin{align}
E_{MN} &=  g_{MN}+ \frac{B_{MN}}{\sqrt{g_s}}, \label{myersdbi2} \\
Q^m{}_n&=\delta^m{}_n \mathds{1}_N + \frac{i}{2\pi} \sqrt{g_s}\, [\phi^m,\phi^l]\, E_{ln}, \\
H_{MN} &= E_{M m} \left(\left(Q^{-1}\right)^m{}_n - \delta^m{}_n \mathds{1}_N\right) E^{nl} E_{l N}. \label{halphabeta}
\end{align}
Here $M,N=0,\ldots,9$ are 10d spacetime indices, $\alpha,\beta=0,\ldots,7$ are indices along the worldvolume of the branes and $m,n=8,9$ are transverse indices. Furthermore, $E^{nl}$ denotes the inverse of $ E_{nl}$.
The two determinants in \eqref{myersdbi} are taken over the $\alpha,\beta$ and $m,n$ indices, respectively. Throughout this paper, we work in string units with $2\pi\sqrt{\alpha'}=1$ and normalize (anti-)commutators such that $[A,B]=AB-BA$, $\{A,B\}=AB+BA$.
Our convention for the norm of an $n$-form $\omega_n$ is $|\omega_n|^2=\frac{1}{n!}\omega_{MN\ldots L}\bar\omega^{MN\ldots L}$.

The scalar fields $\phi^m$ appearing in $Q^m{}_n$ and $\iota^*_\phi$ (see below) parametrize the transverse deformations of the D7-branes.
They are valued in the adjoint of U$(N)$ (before performing the orientifold projection) and thus $N\times N$ matrices. One can also take the interior product of $\phi^m$ with the holomorphic three-form $\Omega$ of the Calabi-Yau manifold to equivalently describe the brane fluctuations in terms of the $(2,0)$-form $\Phi \sim \text{i}_{\phi} \Omega$ \cite{Jockers:2004yj, Jockers:2005zy}, which we will indeed do further below.

The U$(N)$ gauge field $A_\alpha$ is also in the adjoint and thus again an $N\times N$ matrix.\footnote{The orientifold projection projects out some components of $\phi^m$, $A_\alpha$ and breaks the gauge group but this will not be important for our calculation, see \cite[Secs.~2, 7]{Junghans:2026hfm} for a discussion.} We use the convention of \cite{Myers:1999ps} where the gauge-field strength is $F_{\alpha\beta} = \partial_\alpha A_\beta - \partial_\beta A_\alpha + i [A_\alpha, A_\beta]$
and the generators of the gauge group are Hermitian.
We further denote by $\mathcal{D}_\alpha$ the gauge-covariant derivative and by $D_\alpha$ the fully covariant derivative including both the gauge connection and the Levi-Civita one, e.g., $\mathcal{D}_\alpha \phi^m = \partial_\alpha \phi^m + i [A_\alpha, \phi^m]$ and $D_\alpha \phi^m = \partial_\alpha \phi^m + i [A_\alpha, \phi^m] + \Gamma^m_{\alpha n}\phi^n$ (also note that $D_\alpha \phi^\beta = \Gamma^\beta_{\alpha n}\phi^n\neq 0$). We stress that $D_\alpha$ in our notation is \emph{not} the covariant derivative with respect to the worldvolume metric but the $\alpha$ component of the 10d covariant derivative $D_M$ with respect to the ambient-space metric (satisfying $D_M g_{NL}=0$). Accordingly, $\Gamma^m_{\alpha n}$, $\Gamma^\beta_{\alpha n}$, etc.~are components of the 10d Christoffel symbols $\Gamma^M_{NL}$ computed using $g_{MN}$.

The symbol $\iota^*_\phi$ in \eqref{myersdbi} denotes the pull-back onto the D7-brane worldvolume, which depends on the $\phi^m$ fields.
By performing a so-called normal-coordinate expansion,
the pull-back can be written in terms of a pull-back onto the undeformed submanifold at $\phi^m=0$ (denoted by $\iota^*$ in the following) plus a series of higher-order terms organized in powers of $\phi^m$ \cite{Myers:1999ps, Grana:2003ek, Jockers:2004yj}, as we will see in the next section.

Further note that $\text{STr}$ in \eqref{myersdbi} is the symmetrized trace over the gauge indices defined such that one symmetrizes in all orderings of $F_{\alpha\beta}$, $D_\alpha \phi^m$ and $[\phi^m,\phi^n]$
and also in the individual $\phi^m$'s arising in the normal-coordinate expansion we perform below \cite{Tseytlin:1997csa, Myers:1999ps}. This prescription was proposed in \cite{Myers:1999ps} to be reliable until the order $\alpha'^6$ (where factors of $\phi^m$, $A_\alpha$ and derivatives count as $\sqrt{\alpha'}$). This is sufficient for us since the relevant terms for our calculation will be of the order $\alpha'^2$. In fact, all terms at this order turn out to involve only two factors that have to be permuted so that we can from now on replace STr by the ordinary trace (which is cyclic and thus implies the symmetrization at this order). Note that the DBI action also receives curvature corrections at the order $\alpha'^2$ \cite{Bachas:1999um} but they are not relevant for us since they do not affect the $D$-terms \cite{Haack:2006cy}.\footnote{The curvature corrections modify the gauge-kinetic function at the order $\alpha'^2$ \cite{Haack:2006cy}, but this only modifies the $D$-term potential at the order $\alpha'^4$.}

We finally note that we will derive the brane potential under the usual assumption that the backreaction of the localized sources and the fluxes on the 10d bulk spacetime is small so that the dilaton and the warp factor can be treated as approximately constant up to negligible corrections. The 10d metric in this regime is a product of a 4d external metric and a Ricci-flat Calabi-Yau metric, and the D7-brane stack can be treated as a probe on the Calabi-Yau background. This approximation is expected to be valid for sufficiently small string coupling $g_s$ and sufficiently large Calabi-Yau volume $\mathcal{V}$.

\section{Normal-coordinate expansion}
\label{normal}

In order to make the $\phi^m$ dependence of the DBI action \eqref{myersdbi} explicit, we have to evaluate how the pull-back $\iota_\phi^*$ acts on $g_{MN}$, $B_{MN}$ and $H_{MN}$.
For a single brane, we define $\phi^m$ such that the brane is extended along the $x^\alpha$ coordinates and localized at some small $x^m=\frac{1}{2\pi}\phi^m(x^\alpha)$. We then pull back the ambient-spacetime tensors onto the worldvolume with the standard formula and perform a Taylor expansion in small $\phi^m$. The prescription for $N$ branes is \cite{Myers:1999ps} to promote $\phi^m$ to a field in the adjoint of the gauge group and the $\partial_\alpha$ derivatives to gauge-covariant ones.

We first consider the metric. For a single brane, pulling back the metric and expanding up to the second order in $x^m=\frac{1}{2\pi}\phi^m$ yields
\begin{align}
(\iota_\phi^* g)_{\alpha\beta} &= g_{\alpha\beta}(\phi) + \frac{1}{2\pi}g_{\alpha m}(\phi) \partial_\beta \phi^m + \frac{1}{2\pi} g_{m \beta}(\phi) \partial_\alpha \phi^m + \frac{1}{4\pi^2} g_{mn}(\phi) \partial_\alpha \phi^m \partial_\beta \phi^n \nll = 
g_{\alpha\beta}|_0 + \frac{1}{2\pi}\phi^m \partial_m g_{\alpha\beta}|_0 + \frac{1}{2\pi} g_{m \beta}|_0 \partial_\alpha \phi^m  + \frac{1}{2\pi}g_{\alpha m}|_0 \partial_\beta \phi^m \nl
+ \frac{1}{8\pi^2}\phi^m \phi^n \partial_m \partial_n g_{\alpha\beta}|_0
+ \frac{1}{4\pi^2} \phi^n \partial_n g_{m \beta}|_0 \partial_\alpha \phi^m 
+ \frac{1}{4\pi^2} \phi^n \partial_n g_{\alpha m}|_0 \partial_\beta \phi^m \nl
+ \frac{1}{4\pi^2} g_{mn}|_0 \partial_\alpha \phi^m \partial_\beta \phi^n + \mathcal{O}(\alpha'^3), \label{sdsfkusfdaijgsdgh}
\end{align}
where by $|_0$ we mean evaluation at $x^m=0$. Recall that each $\phi^m$ factor and each derivative comes with an implicit factor of $\sqrt{\alpha'}$ so that the terms we wrote down are of the order $\alpha'^2$ or lower.
As we explained, the corresponding non-Abelian expression is obtained by replacing $\partial_\alpha\to\mathcal{D}_\alpha$.

Analogously, we obtain
\begin{align}
(\iota_\phi^* B)_{\alpha\beta} &= B_{\alpha\beta}(\phi) + \frac{1}{2\pi}B_{\alpha m}(\phi) \mathcal{D}_\beta \phi^m  + \frac{1}{2\pi} B_{m \beta}(\phi) \mathcal{D}_\alpha \phi^m + \frac{1}{4\pi^2} B_{mn}(\phi) \mathcal{D}_\alpha \phi^m \mathcal{D}_\beta \phi^n
\nll =
B_{\alpha\beta}|_0 + \frac{1}{2\pi} \phi^m \partial_m B_{\alpha\beta}|_0 
+ \frac{1}{2\pi}B_{\alpha m}|_0 \mathcal{D}_\beta \phi^m  + \frac{1}{2\pi} B_{m \beta}|_0 \mathcal{D}_\alpha \phi^m  + \mathcal{O}(\alpha'^2) \nll =
B_{\alpha\beta}|_0 + \frac{1}{2\pi} \phi^m H_{m \alpha\beta}|_0 + \frac{1}{2\pi}\mathcal{D}_\beta (B_{\alpha m}|_0 \phi^m)  + \frac{1}{2\pi} \mathcal{D}_\alpha (B_{m \beta}|_0 \phi^m)  + \mathcal{O}(\alpha'^2) \label{dfgafgiafelf}
\end{align}
with $\partial_{[m} B_{\alpha\beta]}|_0 = 2 H_{m \alpha\beta}|_0$.\footnote{In our conventions, $\partial_{[m} B_{\alpha\beta]}= 2\partial_{m} B_{\alpha\beta} - 2\partial_{\alpha} B_{m \beta} - 2\partial_{\beta} B_{\alpha m}$.}
Here we only displayed terms until the order $\alpha'$ since the higher-order terms will not be required below. We furthermore have
\begin{align}
(\iota_\phi^* H)_{\alpha\beta} &= \left(E_{\alpha m}(\phi) + \frac{1}{2\pi} E_{n m}(\phi) \mathcal{D}_\alpha \phi^n \right)\!\! \left( - \frac{i\sqrt{g_s}}{2\pi} [\phi^m,\phi^l] - \frac{g_s}{4\pi^2} [\phi^m,\phi^q]\, E_{qp}(\phi) [\phi^p,\phi^l] + \ldots \right)\nl \times \left(E_{l \beta}(\phi) + \frac{1}{2\pi} E_{l r}(\phi) \mathcal{D}_\beta \phi^r\right) \nll
=
- \frac{i \sqrt{g_s} }{2\pi} E_{\alpha m}|_0 [\phi^m,\phi^l] E_{l \beta}|_0
- \frac{i \sqrt{g_s}}{4\pi^2} \left(\phi^n \partial_n E_{\alpha m}|_0 + E_{n m}|_0 \mathcal{D}_\alpha \phi^n \right) [\phi^m,\phi^l] E_{l \beta}|_0 \nl
- \frac{i\sqrt{g_s}}{4\pi^2} E_{\alpha m}|_0 [\phi^m,\phi^l] \left( \phi^r \partial_r E_{l \beta}|_0 + E_{l r}|_0 \mathcal{D}_\beta \phi^r\right) \nl
- \frac{g_s}{4\pi^2} E_{\alpha m}|_0 [\phi^m,\phi^q]\, E_{qp}|_0 [\phi^p,\phi^l] E_{l \beta}|_0 + \mathcal{O}(\alpha'^3). \label{admfadgfjdagf}
\end{align}
Note that terms $\sim \phi^n [\phi^m,\phi^l]$ vanish when taking the trace so that they do not contribute to the action at the order $\alpha'^2$. We will therefore ignore these terms in the following.

In order to simplify \eqref{dfgafgiafelf}, \eqref{admfadgfjdagf}, it is convenient to make a gauge transformation $B_2 \to B_2 + \d \eta$ with $\eta = B_{\alpha m}|_0 x^m \d x^\alpha - \frac{1}{2} B_{mn}|_0 x^m \d x^n$, which sets $B_{\alpha m}|_0=B_{mn}|_0=0$ everywhere on the branes and leaves $B_{\alpha \beta}|_0$ invariant \cite{Menet:2025mbi}.
This yields
\begin{align}
(\iota_\phi^* B)_{\alpha\beta} &= B_{\alpha\beta}|_0 + \frac{1}{2\pi} \phi^m H_{m \alpha\beta}|_0 + \mathcal{O}(\alpha'^2), \\
(\iota_\phi^* H)_{\alpha\beta} &=
- \frac{i \sqrt{g_s} }{2\pi} g_{\alpha m} [\phi^m,\phi^l] g_{l \beta}|_0
- \frac{i \sqrt{g_s}}{4\pi^2} g_{n m} \mathcal{D}_\alpha \phi^n [\phi^m,\phi^l] g_{l \beta}|_0 \nl
- \frac{i\sqrt{g_s}}{4\pi^2} g_{\alpha m} [\phi^m,\phi^l] g_{l r}|_0 \mathcal{D}_\beta \phi^r  
- \frac{g_s}{4\pi^2} g_{\alpha m} [\phi^m,\phi^q]\, g_{qp} [\phi^p,\phi^l] g_{l \beta}|_0 + \ldots \nl + \mathcal{O}(\alpha'^3), \label{fhsfhgsfhhhhhhh}
\end{align}
where the dots stand for ignored terms $\sim \phi^n [\phi^m,\phi^l]$ as explained above.

We now specialize to the case where the 10d spacetime is a product of a 4d external spacetime and a Calabi-Yau 3-fold $X$ (more precisely, an orientifold thereof) and the submanifold wrapped by the D7-branes is a divisor $D$ on $X$. We can then choose local complex coordinates $z^I=\{z^i,z\}$ on $X$ such that the divisor sits at $z=0$. We furthermore denote the (real) coordinates on the external 4d spacetime by $x^\mu$.
Hence, the worldvolume coordinates are $\{x^\mu, z^i, \bar z^{\bar\jmath}\}$ and the transverse ones are $\{z,\bar z\}$. Accordingly, we will write $\phi^z = \phi^8 + i\phi^9$, $\phi^{\bar z} = (\phi^z)^\dagger = \phi^8 - i\phi^9$.

Since Calabi-Yau manifolds are K\"ahler, the non-vanishing metric components are $g_{I \bar J} = \partial_{I}\partial_{\bar J} k$ for some function $k(z^I,\bar z^{\bar J})$. Note that this implies $\partial_K g_{I \bar J} = \partial_I g_{K \bar J}$. The non-vanishing Christoffel symbols are $\Gamma^I_{JK} = g^{I\bar J} \partial_J g_{\bar J K}$ and its complex conjugate $\Gamma^{\bar I}_{\bar J\bar K}$, and the Riemann tensor is
\begin{equation}
\mathcal{R}_{I \bar J K \bar L} = \mathcal{R}_{I \bar L K \bar J} = \mathcal{R}_{K \bar J I \bar L} = \partial_K\partial_{\bar J} g_{I \bar L} - g^{P \bar Q} (\partial_K g_{I \bar Q}) (\partial_{\bar J} g_{\bar L P}). \label{aslkfdif}
\end{equation}

On real manifolds, \eqref{sdsfkusfdaijgsdgh} and \eqref{fhsfhgsfhhhhhhh} can be further simplified by going to so-called Fermi normal coordinates satisfying $g_{\alpha m}|_0=0$, $g_{mn}|_0=\delta_{mn}$, $\partial_l g_{mn}|_0=0$ \cite[Prop.~5.26]{Lee}. See, e.g., \cite{Menet:2025mbi} for a recent application. Unfortunately, there is in general no complex analogue of such coordinates. Indeed, having $g_{i \bar z}|_0=0$, $g_{z \bar z}|_0=\text{const.}$, $\partial_{\bar z} g_{z \bar z}|_0=0$ everywhere on $D$ would imply, e.g., $\partial_{\bar\jmath} g_{i \bar z}|_0= \partial_{\bar\jmath} g_{z \bar z}|_0 = \partial_k \partial_{\bar\jmath} g_{i \bar z}|_0 = \partial_i \partial_{\bar\jmath} g_{z \bar z}|_0 = \partial_i \partial_{\bar z} g_{z \bar z}|_0 = 0$ and, because of $g_{I \bar J} = \partial_{I}\partial_{\bar J} k$, also that $\partial_{\bar z} g_{i \bar \jmath}|_0= \partial_{\bar z} g_{z \bar \jmath}|_0 = 0$, etc. Using \eqref{aslkfdif}, we would then conclude that $\mathcal{R}_{i\bar\jmath z \bar z}|_0 = \mathcal{R}_{i\bar z z \bar z}|_0 = \mathcal{R}_{i \bar\jmath k \bar z}|_0=0$. Since the Calabi-Yau is Ricci-flat, it follows that $\mathcal{R}_{i\bar\jmath k \bar l}g^{k\bar l}|_0 = 0$, which in turn would imply trivial tangent and normal bundles of the divisor (recall that, on a Calabi-Yau, $c_1(ND)=-c_1(TD)$). Hence, we can in general not choose coordinates such that $g_{i \bar z}|_0=0$, $g_{\bar z z}|_0=\text{const.}$, $\partial_{\bar z} g_{z \bar z}|_0=0$ on the brane worldvolume and will therefore keep these metric components arbitrary in the following. This yields rather lengthy expressions in the DBI action, but we will see that the final result is nevertheless very simple.

We first write the different components of the pulled-back metric \eqref{sdsfkusfdaijgsdgh} in a manifestly covariant form:
\begin{align}
(\iota_\phi^* g)_{\mu\nu} &= g_{\mu\nu}|_0 + \frac{1}{4\pi^2} g_{z \bar z }|_0 D_\mu \phi^z D_\nu \phi^{\bar z} 
+ \frac{1}{4\pi^2} g_{z \bar z }|_0 D_\nu \phi^z D_\mu \phi^{\bar z} + \mathcal{O}(\alpha'^3), \\
(\iota_\phi^* g)_{\mu i} &= \frac{1}{2\pi} g_{\bar z i}|_0 D_\mu \phi^{\bar z} + \mathcal{O}(\alpha'^2), \\
(\iota_\phi^* g)_{i\bar\jmath} &= g_{i\bar\jmath}|_0 + \frac{1}{2\pi} D_i (g_{I \bar\jmath}  \phi^I ) |_0  + \frac{1}{2\pi} D_{\bar\jmath} (g_{i \bar J} \phi^{\bar J}) |_0 \nl
+ \frac{1}{8\pi^2} D_i (\phi^J g_{I \bar\jmath} D_J \phi^I )|_0 + \frac{1}{8\pi^2} D_{\bar\jmath} (\phi^{\bar J} g_{\bar I i} D_{\bar J} \phi^{\bar I} )|_0
+ \frac{1}{4\pi^2}\phi^z \phi^{\bar z} \mathcal{R}_{i \bar \jmath z \bar z}|_0 \nl
+ \frac{1}{4\pi^2} g_{I \bar J} D_i \phi^I D_{\bar\jmath} \phi^{\bar J} |_0
+ \frac{1}{4\pi^2} g_{z \bar z}|_0 D_i \phi^{\bar z} D_{\bar\jmath} \phi^{z}
+ \mathcal{O}(\alpha'^3), \label{dgsgsdgsfhhz} \\
(\iota_\phi^* g)_{ij} &= \frac{1}{2\pi} g_{\bar z j}|_0 D_i \phi^{\bar z}  + \frac{1}{2\pi}g_{i \bar z}|_0 D_j \phi^{\bar z} + \mathcal{O}(\alpha'^2). \label{faeoefrkdfaj9dj}
\end{align}
Here we used that $\mathcal{D}_\mu \phi^z=D_\mu \phi^z$, $\mathcal{D}_i \phi^{\bar z}=D_i \phi^{\bar z}$ (since $\Gamma^{z}_{\mu z} = \Gamma^{z}_{\mu \bar z} = \Gamma^{\bar z}_{i z}=\Gamma^{\bar z}_{i \bar z}=0$). Note that $D_i \phi^I$ in \eqref{dgsgsdgsfhhz} is not the same as $D_i \phi^z$ since $D_i \phi^j = \Gamma^j_{iz}\phi^z\neq 0$. For the components $(\iota_\phi^* g)_{\mu i}$ and $(\iota_\phi^* g)_{ij}$, we only displayed terms up to the order $\alpha'$ since we will not need the higher-order terms.

The components of $(\iota_\phi^* H)_{\alpha\beta}$ are
\begin{align}
(\iota_\phi^* H)_{\mu \nu} &= 0 + \mathcal{O}(\alpha'^3), \\
(\iota_\phi^* H)_{\mu i} &= (\iota_\phi^* H)_{ij} = 0 + \mathcal{O}(\alpha'^2), \\
(\iota_\phi^* H)_{i\bar\jmath} &= i \frac{\sqrt{g_s}}{2\pi} [\phi^z,\phi^{\bar z}] g_{i\bar z} g_{z \bar\jmath}|_0
+i \frac{\sqrt{g_s}}{4\pi^2} (\mathcal{D}_i \phi^z ) [\phi^z,\phi^{\bar z}] g_{z \bar z} g_{z \bar\jmath}|_0 \nl
+ i \frac{\sqrt{g_s}}{4\pi^2} [\phi^z,\phi^{\bar z}] (\mathcal{D}_{\bar\jmath}\phi^{\bar z}) g_{i\bar z} g_{z \bar z} |_0
- \frac{g_s}{4\pi^2} [\phi^z,\phi^{\bar z}]^2 g_{i\bar z} g_{z \bar z} g_{z \bar\jmath}|_0 + \mathcal{O}(\alpha'^3). \label{dsgfsdgfsdjlgf}
\end{align}
We again only displayed $\mathcal{O}(\alpha'^2)$ terms for those components for which we will need them below.

\section{Rewriting the action}
\label{exp}

We are now ready to expand the DBI action \eqref{myersdbi} up to the order $\alpha'^2$. The second determinant in \eqref{myersdbi} yields
\begin{equation}
\sqrt{\text{det}(Q^m{}_n)} = \mathds{1}_N + \frac{g_s}{8\pi^2} g_{z\bar z}^2[\phi^z,\phi^{\bar z}]^2|_0 + \mathcal{O}(\alpha'^3). \label{dsfsdofgds}
\end{equation}
Here we again ignored cubic terms $\sim \phi^n [\phi^m,\phi^l]$ since they vanish under the trace in the action.
To expand the first determinant, we use the formula
\begin{equation}
\sqrt{-\text{det} (g_{\alpha\beta}|_0 \mathds{1}_N + h_{\alpha\beta})} = \sqrt{-\text{det} (g_{\alpha\beta}|_0)} \left(\mathds{1}_N + \frac{1}{2} h_\alpha^\alpha
+ \frac{1}{8} (h_\alpha^\alpha)^2 - \frac{1}{4} h_{\alpha\beta} h^{\beta\alpha} + \mathcal{O}(h^3) \right) \label{dsjgisglns}
\end{equation}
with $h_{\alpha\beta} = (\iota^*_\phi g)_{\alpha\beta} - g_{\alpha\beta}|_0 \mathds{1}_N + \frac{1}{\sqrt{g_s}}(\iota^*_\phi B)_{\alpha\beta} + (\iota^*_\phi H)_{\alpha\beta} + \frac{F_{\alpha\beta}}{2\pi \sqrt{g_s}}$.
Here and in the following, indices are raised with the inverse of $g_{\alpha\beta}|_0$, which we denote by $\hat g^{\alpha\beta}|_0$. Note that $\hat g^{\alpha\beta}|_0$ is not the same as $g^{\alpha\beta}|_0$ (the $\alpha,\beta$ component of $g^{MN}|_0$). From $\delta^\alpha_\beta= \hat g^{\alpha \gamma} g_{\gamma \beta } = g^{\alpha M} g_{M \beta}$, one can derive the formulae
\begin{equation}
g^{i \bar z} = - \hat g^{i \bar \jmath} g^{z \bar z} g_{z \bar \jmath},
\qquad \hat  g^{i\bar \jmath}g^{z \bar z} = g^{i \bar \jmath} g^{z \bar z} - g^{i\bar z} g^{z \bar \jmath},
\qquad \left( g_{z \bar z}- \hat g^{i\bar\jmath} g_{i \bar z} g_{z \bar\jmath} \right) g^{z \bar z} = 1, \label{dsgsldigsdljg}
\end{equation}
which will become useful further below.

Using \eqref{dsfsdofgds}, \eqref{dsjgisglns} and the expressions for $(\iota^*_\phi g)_{\alpha\beta}$, $(\iota^*_\phi B)_{\alpha\beta}$, $(\iota^*_\phi H)_{\alpha\beta}$ derived in the previous section, the integrand of the DBI action becomes
\begin{align}
\mathcal{L}_\text{DBI} &= -2\pi g_s \text{Tr} \sqrt{-\text{det}\left( (\iota^*_\phi E)_{\alpha\beta} + (\iota^*_\phi H)_{\alpha\beta} + \frac{F_{\alpha\beta}}{2\pi \sqrt{g_s}} \right)} \sqrt{\text{det}(Q^m{}_n)} \nll
 = -2\pi g_s \text{Tr} \sqrt{-\text{det} g_{\alpha\beta}|_0} \Bigg[ \mathds{1}_N
+ \frac{1}{4\pi^2} \left( g_{z \bar z } - \hat g^{i\bar\jmath} g_{\bar z i} g_{z \bar\jmath} \right)\!|_0 D_\mu \phi^z D^\mu \phi^{\bar z} \nl
+ \frac{1}{8\pi^2} \left[ D_i (g_{I \bar\jmath}  \phi^I )  + D_{\bar\jmath} (g_{i \bar J} \phi^{\bar J}) \right] \! \left( \hat g^{i\bar\jmath} \hat g^{k \bar l} -  \hat g^{i\bar l} \hat g^{k\bar\jmath}\right) \!
\left.\left[ D_k (g_{K \bar l}  \phi^K ) + D_{\bar l} (g_{k \bar L} \phi^{\bar L}) \right]\right|_0 \nl
+ \frac{1}{4\pi^2}\phi^z \phi^{\bar z} \hat g^{i\bar\jmath} \mathcal{R}_{i \bar \jmath z \bar z}|_0 
+ \frac{1}{4\pi^2} g_{I \bar J} \hat g^{i\bar\jmath} D_i \phi^I D_{\bar\jmath} \phi^{\bar J} |_0
+ \frac{1}{4\pi^2} g_{z \bar z} \hat g^{i\bar\jmath} D_i \phi^{\bar z} D_{\bar\jmath} \phi^{z} |_0 \nl
- \frac{1}{8\pi^2} \left( g_{\bar z j} D_i \phi^{\bar z} + g_{i \bar z} D_j \phi^{\bar z} \right) \hat g^{i \bar l} \hat g^{j \bar k} \left( g_{ z \bar l } D_{\bar k} \phi^{z} + g_{\bar k z} D_{\bar l}\phi^{z} \right)|_0
+ \frac{1}{4g_s} \mathcal{F}_{\mu\nu}\mathcal{F}^{\mu\nu} \nl
+ \frac{1}{2g_s}\left|\mathcal{F}_{i \bar\jmath} + \frac{1}{2\pi}\phi^z H_{i\bar\jmath z}|_0 + \frac{1}{2\pi}\phi^{\bar z} H_{i\bar\jmath \bar z}|_0 \right|^2  + \frac{1}{2g_s}\left|\mathcal{F}_{i j} + \frac{1}{2\pi}\phi^z H_{ijz}|_0 + \frac{1}{2\pi}\phi^{\bar z} H_{ij \bar z}|_0\right|^2  \nl
+ \frac{1}{g_s} \mathcal{F}_{\mu i}\mathcal{F}^{\mu i} 
-  \frac{i}{2\pi} [\phi^z,\phi^{\bar z}] g_{i\bar z} g_{z \bar\jmath}|_0 \mathcal{F}^{\bar\jmath i}
+ \frac{g_s}{8\pi^2} \left( g_{z\bar z} - \hat g^{i\bar\jmath} g_{i\bar z} g_{z \bar\jmath}\right)^2|_0 [\phi^z,\phi^{\bar z}]^2 \Bigg] \nl + \text{t.d.} +
\mathcal{O}(\alpha'^3). \label{asfsafdtdgtt}
\end{align}
Note that we combined $B_{\alpha\beta}|_0$ and $F_{\alpha\beta}$ into the gauge-invariant field strength $\mathcal{F}_{\alpha\beta}= B_{\alpha\beta}|_0 \mathds{1}_N + \frac{1}{2\pi}F_{\alpha\beta}$. Here and in the following, each factor of $\mathcal{F}$ will be counted as being of the order $\alpha'$ in our expansion. Our convention for the square terms is $|\omega_{i\bar\jmath}|^2 = \omega_{i\bar\jmath} \omega_{\bar k l}\hat g^{i\bar k}\hat g^{l \bar\jmath}|_0$, $|\omega_{i j}|^2 = \omega_{i j} \omega_{\bar k \bar l}\hat g^{i\bar k}\hat g^{j \bar l}|_0$.

To simplify \eqref{asfsafdtdgtt}, we integrate by parts and use
\begin{align}
& \text{Tr} (\phi^z D_{[i} D_{\bar \jmath]}\phi^{\bar z}) = \text{Tr} (\mathcal{R}_{i \bar \jmath}{}^{\bar z}{}_{\bar z} \phi^z \phi^{\bar z} - i F_{i\bar \jmath} [\phi^z,\phi^{\bar z}] ) = \text{Tr} (\mathcal{R}_{i \bar \jmath}{}^{\bar z}{}_{\bar z} \phi^z \phi^{\bar z} - 2\pi i \mathcal{F}_{i\bar \jmath} [\phi^z,\phi^{\bar z}] ), \\
& \text{Tr} (\phi^z D_{[i} D_{\bar \jmath]}\phi^{\bar k}) = \text{Tr} (\mathcal{R}_{i \bar \jmath}{}^{\bar k}{}_{\bar z} \phi^z \phi^{\bar z}).
\end{align}
This yields
\begin{align}
\mathcal{L}_\text{DBI} &= -2\pi g_s \text{Tr} \sqrt{-\text{det} g_{\alpha\beta}|_0} \Bigg[ \mathds{1}_N
+ \frac{1}{4\pi^2} \left( g_{z \bar z } - \hat g^{i\bar\jmath} g_{\bar z i} g_{z \bar\jmath} \right) D_\mu \phi^z D^\mu \phi^{\bar z} |_0 + \frac{1}{4g_s} \mathcal{F}_{\mu\nu}\mathcal{F}^{\mu\nu} \nl
+ \frac{i}{2\pi} \left( g_{z \bar z}- \hat g^{k \bar l} g_{\bar z k} g_{z \bar l} \right) \hat g^{i\bar\jmath} \mathcal{F}_{i \bar\jmath} [\phi^z ,\phi^{\bar z}] |_0 
+ \frac{1}{2\pi^2} \left( g_{z \bar z} - \hat g^{k \bar l} g_{\bar z k} g_{z \bar l}\right) \hat g^{i\bar\jmath} D_i \phi^{\bar z} D_{\bar\jmath} \phi^{z} |_0 \nl
+ \frac{1}{2g_s}\left|\mathcal{F}_{i \bar\jmath} + \frac{1}{2\pi}\phi^z H_{i\bar\jmath z}|_0 + \frac{1}{2\pi}\phi^{\bar z} H_{i\bar\jmath \bar z}|_0 \right|^2  + \frac{1}{2g_s}\left|\mathcal{F}_{i j} + \frac{1}{2\pi}\phi^z H_{ijz}|_0 + \frac{1}{2\pi}\phi^{\bar z} H_{ij \bar z}|_0\right|^2  \nl
+ \frac{1}{g_s} \mathcal{F}_{\mu i}\mathcal{F}^{\mu i}
+ \frac{g_s}{8\pi^2} \left( g_{z\bar z} - \hat g^{i\bar\jmath} g_{i\bar z} g_{z \bar\jmath}\right)^2 [\phi^z,\phi^{\bar z}]^2 |_0  \Bigg] + \text{t.d.} +
\mathcal{O}(\alpha'^3). \label{dfadifadlijf}
\end{align}
Note that all curvature terms $\sim \mathcal{R}\phi^2$ have cancelled out.

Let us now rewrite the two squares in the third line, where we abbreviate $(\mathcal{F}_\phi)_{\alpha\beta} := \mathcal{F}_{\alpha\beta}+\frac{1}{2\pi}\phi^mH_{\alpha\beta m}|_0$, $\mathcal{F}_\phi= \frac{1}{2}(\mathcal{F}_\phi)_{\alpha\beta}\d x^\alpha \w \d x^\beta$ for convenience. Hence, $\mathcal{F}_\phi$ corresponds to the worldvolume flux on the fluctuating brane worldvolume at $\phi^m\neq 0$, while $\mathcal{F}$ denotes the worldvolume flux at $\phi^m=0$.
With this notation, the squares in the third line of \eqref{dfadifadlijf} are simply $\frac{1}{2g_s}|(\mathcal{F}_\phi)_{i\bar\jmath} |^2  + \frac{1}{2g_s}|(\mathcal{F}_\phi)_{ij}|^2$.

In general, any 2-form $f$ on the divisor can always be split into its $(1,1)$, $(2,0)$ and $(0,2)$ parts, and one can show that they satisfy
\begin{equation}
\star |f_{i\bar\jmath}|^2 = \star |\iota^* J\cdot f|^2 -  f^{1,1} \w f^{1,1}, \qquad \star |f_{i j}|^2 = 2 f^{2,0} \w f^{0,2},
\end{equation}
where $|\iota^* J\cdot f|^2= \left[(\iota^* J)^{i\bar\jmath}f_{i\bar\jmath}\right]^2$, $\iota^* J$ is the K\"ahler form on $D$ inherited from the one on $X$ (satisfying $(\iota^* J)_{i\bar\jmath} = -i g_{i\bar\jmath}|_0$) and by $\star$ we denote the Hodge operator on $D$, which satisfies $\star 1 = \frac{1}{2} \iota^* J \w \iota^* J $ in our conventions.
Hence,
\begin{equation}
|f_{i\bar\jmath}|^2 + |f_{i j}|^2 = |\iota^* J\cdot f|^2 + 2 |f_{i j}|^2 - \star (f\w f).
\end{equation}
Since we consider an ISD flux background, we further have that the bulk fluxes are primitive, i.e., $J^{I\bar J} H_{I\bar J K} =0$ \cite{Giddings:2001yu}, where the indices are raised with the Calabi-Yau metric.
One can check using \eqref{dsgsldigsdljg} that this implies $(\iota^* J)^{i\bar\jmath} H_{i\bar\jmath z}|_0 =0$ (where the indices are now raised with $\hat g^{i\bar\jmath}|_0$ and we assumed $H_{i\bar \jmath k}|_0=0$ to avoid a Freed-Witten anomaly \cite{Freed:1999vc}) and therefore $|\iota^* J\cdot \mathcal{F}_\phi|^2 = |\iota^* J\cdot \mathcal{F}|^2$. Using this in the third line of \eqref{dfadifadlijf}, we obtain
\begin{align}
\mathcal{L}_\text{DBI} &= -2\pi g_s \text{Tr} \sqrt{-\text{det} g_{\alpha\beta}|_0} \Bigg[ \mathds{1}_N
+ \frac{1}{4\pi^2} \left( g_{z \bar z } - \hat g^{i\bar\jmath} g_{\bar z i} g_{z \bar\jmath} \right) D_\mu \phi^z D^\mu \phi^{\bar z} |_0 \nl + \frac{1}{2\pi^2} \left( g_{z \bar z} - \hat g^{k \bar l} g_{\bar z k} g_{z \bar l}\right) \hat g^{i\bar\jmath} D_i \phi^{\bar z} D_{\bar\jmath} \phi^{z} |_0  + \frac{1}{4g_s} \mathcal{F}_{\mu\nu}\mathcal{F}^{\mu\nu} + \frac{1}{g_s} \mathcal{F}_{\mu i}\mathcal{F}^{\mu i}
\nl
+ \frac{1}{2g_s}\left|\iota^* J \cdot \mathcal{F}\right|^2  + \frac{1}{g_s}\left|\mathcal{F}_{i j} + \frac{1}{2\pi}\phi^z H_{ijz}|_0 + \frac{1}{2\pi}\phi^{\bar z} H_{ij \bar z}|_0\right|^2 - \frac{1}{2g_s} \star (\mathcal{F}_\phi \w \mathcal{F}_\phi) \nl
+ \frac{i}{2\pi} \left( g_{z \bar z}- \hat g^{k \bar l} g_{\bar z k} g_{z \bar l} \right) \hat g^{i\bar\jmath} \mathcal{F}_{i \bar\jmath}[\phi^z , \phi^{\bar z}] |_0 + \frac{g_s}{8\pi^2} \left( g_{z\bar z} - \hat g^{i\bar\jmath} g_{i\bar z} g_{z \bar\jmath}\right)^2 [\phi^z,\phi^{\bar z}]^2 |_0 \Bigg] \nl + \text{t.d.} +
\mathcal{O}(\alpha'^3). \label{dgdaoppekfkfdfd}
\end{align}

An important point is now that the first term in the first line of \eqref{dgdaoppekfkfdfd} cancels with the tensions of the other D7-branes and O7-planes in the compactification due to the D7-tadpole condition.
Similarly, the $\mathcal{F}_\phi \w \mathcal{F}_\phi$ term in the third line is canceled due to the D3-tadpole condition, which implies an exactly opposite contribution from other D3/D7-branes, O3/O7-planes and ISD bulk fluxes.
Imposing that the D3 and D7 tadpole conditions are satisfied thus amounts to adding a term
\begin{equation}
\mathcal{L}_\text{tadpoles} = -2\pi g_s \text{Tr} \sqrt{-\text{det} g_{\alpha\beta}|_0} \left[ -\mathds{1}_N + \frac{1}{2g_s}\star \left(\mathcal{F}_\phi \w \mathcal{F}_\phi \right) \right].
\end{equation}
Note that, in general, the D7-branes affect the bulk solution in a more complicated way than just by contributing to the tadpole, e.g., by sourcing a varying profile for the axio-dilaton, the warp factor, and other fields. However, these backreaction effects are expected to be negligible in the small-$g_s$, large-volume regime we consider compared to the leading brane potential.

From now on, we will also express the bulk flux $H_3$ in terms of $G_3$ \cite{Giddings:2001yu}, as this is the more common notation in the literature.
Using $H_3 =  \frac{i g_s}{2} \left(G_3 - \bar G_3\right)$ and that ISD flux satisfying $iG_3=\star_6 G_3$ can only have $(0,3)$ and $(2,1)$ components\footnote{This is true in the convention of \cite{Giddings:2001yu} where the complex basis is normalized as $\star_6 1 = - i\, \text{det}(g_{I\bar J}) \d z^1\w\d z^2\w \d z^3 \w \d {\bar z}^{\bar 1}\w \d {\bar z}^{\bar 2}\w \d {\bar z}^{\bar 3}$.}, we identify
\begin{equation}
H_{i j z} = - \frac{i g_s}{2} \bar G_{i j z},\qquad H_{i j \bar z} = \frac{i g_s}{2} G_{i j \bar z}.
\end{equation}
Using this in \eqref{dgdaoppekfkfdfd} and adding the tadpole term as explained, we get
\begin{align}
\mathcal{L}_\text{DBI} + \mathcal{L}_\text{tadpoles} &= -2\pi g_s \text{Tr} \sqrt{-\text{det} g_{\alpha\beta}|_0} \Bigg[ \frac{1}{4\pi^2} \left( g_{z \bar z } - \hat g^{i\bar\jmath} g_{\bar z i} g_{z \bar\jmath} \right) D_\mu \phi^z D^\mu \phi^{\bar z} |_0 \nl + \frac{1}{2\pi^2} \left( g_{z \bar z} - \hat g^{k \bar l} g_{\bar z k} g_{z \bar l}\right) \hat g^{i\bar\jmath} D_i \phi^{\bar z} D_{\bar\jmath} \phi^{z} |_0  + \frac{1}{4g_s} \mathcal{F}_{\mu\nu}\mathcal{F}^{\mu\nu}  + \frac{1}{g_s} \mathcal{F}_{\mu i}\mathcal{F}^{\mu i} 
\nl
+ \frac{1}{2g_s}\left|\iota^* J \cdot \mathcal{F}\right|^2  + \frac{1}{g_s}\left|\mathcal{F}_{i j} - \frac{i g_s}{4\pi} \phi^z \bar G_{ijz}|_0 + \frac{i g_s}{4\pi} \phi^{\bar z} G_{ij \bar z}|_0\right|^2 \nl
+ \frac{i}{2\pi} \left( g_{z \bar z}- \hat g^{k \bar l} g_{\bar z k} g_{z \bar l} \right) \hat g^{i\bar\jmath} \mathcal{F}_{i \bar\jmath}[\phi^z , \phi^{\bar z}] |_0 + \frac{g_s}{8\pi^2} \left( g_{z\bar z} - \hat g^{i\bar\jmath} g_{i\bar z} g_{z \bar\jmath}\right)^2 [\phi^z,\phi^{\bar z}]^2 |_0 \Bigg] \nl + \text{t.d.} +
\mathcal{O}(\alpha'^3). \label{jkpojtjkotjkj}
\end{align}

The next step is to express the $\phi^z$ field in terms of a $(2,0)$-form $\Phi$ by contracting with the holomorphic 3-form $\Omega$ of the Calabi-Yau manifold.
To this end, we first note that
\begin{align}
\Omega_{ij z }\bar \Omega_{\bar k \bar l \bar z}\hat g^{i\bar k}\hat g^{j\bar l}g^{z\bar z} &=
\Omega_{ij z }\bar \Omega_{\bar k \bar l \bar z} \left( g^{i\bar k} - \frac{g^{i\bar z} g^{z \bar k}}{g^{z\bar z}}\right) \left( g^{j\bar l} - \frac{g^{j\bar z} g^{z \bar l}}{g^{z\bar z}}\right) g^{z\bar z} \nll
= \Omega_{ij z }\bar \Omega_{\bar k \bar l \bar z} \left( g^{i\bar k} g^{j\bar l} g^{z\bar z} - g^{i\bar z} g^{z \bar k} g^{j\bar l} - g^{i\bar k} g^{j\bar z} g^{z \bar l}\right) \nll
= \frac{1}{3}\Omega_{IJK }\bar \Omega_{\bar I \bar J \bar K} g^{I \bar I} g^{J\bar J} g^{K\bar K} = 2|\Omega|^2,
\end{align}
where we used \eqref{dsgsldigsdljg}.
Note that, on the left-hand side of this equation, we contract with the hatted metric, while we contract with the unhatted metric on the right-hand side. Since $\Omega_{IJK}$ is covariantly constant on a Calabi-Yau manifold, $|\Omega|^2$ and therefore also 
$\Omega_{ij z }\bar \Omega_{\bar k \bar l \bar z}\hat g^{i\bar k}\hat g^{j\bar l}g^{z\bar z}$ must be constant. According to the last identity in \eqref{dsgsldigsdljg}, this implies
\begin{equation}
\Omega_{ij z }\bar \Omega_{\bar k \bar l \bar z}\hat g^{i\bar k}\hat g^{j\bar l} = \frac{\lambda}{g^{z \bar z}} = \lambda \left(g_{z \bar z}- \hat g^{i\bar\jmath} g_{\bar z i} g_{z \bar\jmath}\right) \label{dgsthkotkt}
\end{equation}
for some real constant $\lambda=2|\Omega|^2$ that depends on how we normalize $\Omega$.
Another useful identity is
\begin{equation}
\Omega_{ijz}\bar\Omega^{klz} = |\Omega|^2 \delta^k_{[i} \delta^l_{j]}, \label{fjsifjdslf}
\end{equation}
where indices can equivalently be raised with hatted or unhatted metrics, $\bar\Omega^{klz} = \bar\Omega_{\bar k\bar l\bar z} \hat g^{k\bar k} \hat g^{l\bar l} g^{z\bar z}$ $ =  \bar\Omega_{\bar I\bar J \bar K}  g^{k\bar I} g^{l\bar J} g^{z\bar K}$.
Using \eqref{dgsthkotkt}, \eqref{fjsifjdslf} in \eqref{jkpojtjkotjkj} and performing the field redefinition
\begin{equation}
\Phi_{ij}= \frac{\sqrt{g_s}}{|\Omega|} \Omega_{ij z}|_0\phi^z, \qquad \Phi^\dagger_{\bar\imath\bar\jmath}= \frac{\sqrt{g_s}}{|\Omega|} \bar\Omega_{\bar\imath\bar\jmath \bar z}|_0\phi^{\bar z}
\end{equation}
then yields
\begin{align}
\mathcal{L}_\text{DBI} + \mathcal{L}_\text{tadpoles} &= \text{Tr} \sqrt{-\text{det} g_{\alpha\beta}|_0} \Bigg[
-\frac{1}{4\pi} (D_\mu \Phi_{ij}) (D^\mu \Phi^\dagger{}^{ij})
- \frac{\pi}{2} \mathcal{F}_{\mu\nu}\mathcal{F}^{\mu\nu}  - 2\pi \mathcal{F}_{\mu i}\mathcal{F}^{\mu i} \nl - \frac{1}{2\pi} (D_{\bar\imath} \Phi_{jk})(D^{\bar\imath} \Phi^\dagger{}^{jk}) - \pi \left( |\iota^* J\cdot \mathcal{F}| + \frac{1}{4\pi} [\Phi_{ij},\Phi^\dagger{}^{ij}] \right)^2  \nl - 2 \pi \left|\mathcal{F}_{i j} - \frac{i\sqrt{g_s}}{8\pi|\Omega|}\Phi_{ij} \bar\Omega^{klz} \bar G_{klz} |_0 + \frac{i \sqrt{g_s}}{8\pi |\Omega|} \Phi^\dagger_{\bar k\bar l} \Omega^{\bar k\bar l \bar z} G_{ij \bar z} |_0\right|^2
 \Bigg] \nl + \text{t.d.} + \mathcal{O}(\alpha'^3). \label{dgfaidjgfdjgg}
\end{align}
This is equivalent to \eqref{jkpojtjkotjkj} up to $D_\mu$ derivatives acting on $\frac{\sqrt{g_s}}{|\Omega|} \Omega_{ijz}|_0$. The latter may contribute to the kinetic terms of the bulk moduli in a complete derivation of the bulk+brane effective field theory but can be ignored for our purpose since we only derive the brane potential at fixed bulk moduli here.

Integrating \eqref{dgfaidjgfdjgg} over the divisor $D$, we obtain the effective action at the order $\alpha'^2$
\begin{align}
S_\text{eff} & = \int \text{dvol}_4 \,\text{Tr} \int_D \star\, \bigg(- \frac{1}{4\pi} (D_\mu \Phi_{ij})(D^\mu \Phi^\dagger{}^{ij})
- \frac{\pi}{2} \mathcal{F}_{\mu\nu}\mathcal{F}^{\mu\nu} - 2\pi  \mathcal{F}_{\mu i}\mathcal{F}^{\mu i}  \nl - \frac{1}{2\pi}  (D_{\bar\imath} \Phi_{jk})(D^{\bar\imath} \Phi^\dagger{}^{jk})
 -\pi  \left( |\iota^* J\cdot \mathcal{F}| + \frac{1}{4\pi} [\Phi_{ij},\Phi^\dagger{}^{ij}] \right)^2 \nl
- 2 \pi  \left|\mathcal{F}_{i j} - \frac{i\sqrt{g_s}}{8\pi|\Omega|}\Phi_{ij} \bar\Omega^{klz} \bar G_{klz} |_0 + \frac{i \sqrt{g_s}}{8\pi |\Omega|} \Phi^\dagger_{\bar k\bar l} \Omega^{\bar k\bar l \bar z}  G_{ij \bar z} |_0\right|^2 \bigg), \label{sdgmisldglisd2}
\end{align}
where we denote by $\text{dvol}_4$ the volume form of the external 4d spacetime.

Let us pause for a moment and take stock of what we have done so far. We considered the non-Abelian DBI action on a divisor of a Calabi-Yau orientifold including the effects of worldvolume fluxes and bulk 3-form fluxes and taking into account the global consistency condition that the D7 and D3 tadpoles have to be canceled. We rewrote the action in terms of complex coordinates on $D$ and expressed everything in terms of the background data $\Omega$, $J$, $G_3$ and the brane fields $\Phi_{ij}$, $A_\mu$, $A_i$ (where the dependence on the latter two is implicit in $\mathcal{F}_{\alpha\beta}$ and $D_\alpha$).
In order to obtain from this the 4d $D$-term and $F$-term potentials in the familiar language of 4d $\mathcal{N}=1$ supergravity, we still have to integrate out certain KK modes and convert to the 4d Einstein frame. We will discuss this in the next sections. Nevertheless, the action in its present form already allows us to read off the 8d BPS equations that determine the microscopic dynamics underlying the corresponding 4d vacuum solutions. To this end, we set to zero all terms involving external indices and read off from \eqref{sdgmisldglisd2} that the effective potential is
\begin{align}
V &\propto \text{Tr} \Bigg[ 2 \pi \left|\mathcal{F}_{i j} - \frac{i\sqrt{g_s}}{8\pi |\Omega|}\Phi_{ij} \bar\Omega^{klz} \bar G_{klz} |_0 + \frac{i \sqrt{g_s}}{8\pi |\Omega|}\Phi^\dagger_{\bar k\bar l} \Omega^{\bar k\bar l \bar z } G_{ij \bar z} |_0\right|^2 \nl
+ \frac{1}{2\pi} (D_{\bar\imath} \Phi_{jk})(D^{\bar\imath} \Phi^\dagger{}^{jk}) + \pi \left( |\iota^* J\cdot \mathcal{F}| + \frac{1}{4\pi} [\Phi_{ij}, \Phi^\dagger{}^{ij}] \right)^2  \Bigg]. \label{djififejf}
\end{align}
BPS solutions are obtained by minimizing the three squares:
\begin{align}
& \mathcal{F}_{i j} - \frac{i\sqrt{g_s}}{8 \pi |\Omega|}\Phi_{ij} \bar\Omega^{klz} \bar G_{klz} |_0 + \frac{i \sqrt{g_s}}{8\pi |\Omega|}\Phi^\dagger_{\bar k\bar l} \Omega^{\bar k\bar l \bar z}  G_{ij \bar z} |_0 = 0, \notag \\ & D_{\bar\imath} \Phi_{jk} = 0, \qquad |\iota^* J\cdot \mathcal{F}| + \frac{1}{4\pi} [\Phi_{ij},\Phi^\dagger{}^{ij}] = 0. \label{eaopaseghvbksn}
\end{align}
Note that, setting the bulk fluxes $G_{ij z}$ and $G_{ij \bar z}$ to zero, \eqref{eaopaseghvbksn} reproduces the BPS equations stated in \cite[Sec.~3.3.1]{Beasley:2008dc} (up to conventions). See also, e.g., \cite{Gomis:2005wc, Martucci:2006ij} for discussions of the Abelian brane potential and BPS equations. What we found here is a generalization of these equations which describe how the bulk fluxes backreact on non-Abelian brane solutions such as T-branes.
Varying the potential \eqref{djififejf}, one can furthermore study non-BPS solutions, i.e., minima of \eqref{djififejf} in which the three squares do not individually vanish. An example where this is relevant is in particular the T-brane-uplifting mechanism proposed in \cite{Cicoli:2015ylx}, which can be studied from the 8d point of view using the above expressions.
We leave such an analysis for future work.

For convenience, we note that the action can equivalently be written as
\begin{align}
S_\text{eff} &= \int \text{dvol}_4 \,\text{Tr} \int_D \Bigg[ - \frac{1}{2\pi} (D_\mu \Phi^\dagger)\w(D^\mu \Phi) - \frac{1}{4\pi} \star (D_\mu A )\w (D^\mu A ) \nl - \frac{1}{\pi} \star \left| \bar\partial \Phi + i[A,\Phi] \right|^2 - \pi \star \left|\iota^* J \w \mathcal{F} + \frac{1}{2\pi} [ \Phi, \Phi^\dagger]\right|^2 - 4 \pi \star\big| \mathcal{F}_\phi^{2,0}\big|^2
 \Bigg], \label{faofjffegfe}
\end{align}
where $A=A_i \d z^i+A_{\bar\jmath} \d z^{\bar\jmath}$ denotes the internal part of the gauge field. Here and in the following, we do not display anymore the terms involving the 4d vector field $A_\mu$ in the action, as they are not relevant for the computation of the brane potential. We will however keep the kinetic terms for $\Phi$ and $A$ to keep track of their normalization. Further recall that $\mathcal{F}^{2,0}_\phi = \mathcal{F}^{2,0} - \frac{i\sqrt{g_s}}{8 \pi |\Omega|}\Phi\, \bar\Omega^{klz} \bar G_{klz} |_0 + \frac{i \sqrt{g_s}}{16\pi |\Omega|}\Phi^\dagger_{\bar k\bar l} \Omega^{\bar k\bar l \bar z}  G_{ij \bar z} |_0 \d z^i\w \d z^j$ is the fluctuating worldvolume flux as before. The norms of the gauge-bundle-valued forms are defined analogously to the norm of an ordinary form (replacing complex conjugation by Hermitian conjugation) and all contractions are with respect to $\hat g^{i\bar\jmath}|_0$, e.g., $|\mathcal{F}_\phi^{2,0}|^2 = \frac{1}{2} (\mathcal{F}_\phi^{2,0})_{ij} (\mathcal{F}_\phi^{2,0})^\dagger_{\bar\imath\bar\jmath} \hat g^{i \bar\imath} \hat g^{j\bar\jmath}|_0$.
The BPS equations in this notation are
\begin{equation}
\mathcal{F}^{2,0}_\phi = 0, \qquad \bar\partial \Phi + i[A,\Phi] = 0, \qquad \iota^* J \w \mathcal{F} + \frac{1}{2\pi} [ \Phi, \Phi^\dagger] = 0.
\end{equation}

\section{Expansion in harmonics and discussion of KK corrections}
\label{harmo}

We will see in Section \ref{4dd} that the terms in the second line of \eqref{faofjffegfe} can be identified with the 4d $D$-term and $F$-term potential of the D7-branes. However, in order to match the expressions, we first have to perform an expansion in harmonics and integrate out various KK modes, which is the topic of the present section.

\subsection{Setting up the expansion}

Let us first introduce some notation and recall a few basic facts. Since we study the dynamics in a background with non-trivial gauge bundle (i.e., non-vanishing worldvolume flux), we split $A$ into a background field $\langle A\rangle$ and a fluctuation,
\begin{equation}
A=\langle A \rangle + a + a^\dagger, \label{dfafjilsdglijsg}
\end{equation}
where we define $a$ to be a $(0,1)$-form. We further split $a$ and $\Phi$ into zero and KK modes,
\begin{equation}
a= a_\text{h} + a_\text{KK}, \qquad \Phi = \Phi_\text{h} + \Phi_\text{KK}, \label{dsgsdgsdgkkjkj}
\end{equation}
where $a_\text{h}$ and $\Phi_\text{h}$ are harmonic forms defined with respect to the gauge-covariant derivative operator $\bar\partial_{\langle A \rangle}$, i.e.,
\begin{equation}
\bar\partial_{\langle A \rangle} a_\text{h} = \partial_{\langle A \rangle} \star a_\text{h} = 0, \qquad \bar\partial_{\langle A \rangle} \Phi_\text{h} = \partial_{\langle A \rangle} \star \Phi_\text{h} = 0. \label{dgsgjdsgjig}
\end{equation}
Here $\bar\partial_{\langle A\rangle} = \bar\partial + i[\langle A^{0,1}\rangle,\cdot]$, $\partial_{\langle A\rangle} = \partial + i[\langle A^{1,0}\rangle,\cdot]$ for even-rank forms and $\bar\partial_{\langle A\rangle} = \bar\partial + i\{ \langle A^{0,1}\rangle,\cdot \}$, $\partial_{\langle A\rangle} = \partial + i\{\langle A^{1,0}\rangle,\cdot\}$ for odd-rank forms. Hence, $a_\text{h}$, $\Phi_\text{h}$ contain the zero modes of $a$, $\Phi$, whereas $a_\text{KK}$, $\Phi_\text{KK}$ contain all non-harmonic fluctuations and thus correspond to the KK modes.

Recall that a bundle-valued $(0,q)$-form $\omega_\text{h}$ is $\bar\partial_{\langle A\rangle}$-harmonic iff it is $\bar\partial_{\langle A\rangle}$-closed and $\partial_{\langle A\rangle}$-co-closed. Such forms are classified by the cohomology groups $H^q(D,V)$, where $V$ is the gauge bundle. By the Hodge decomposition, we can furthermore write any $(0,q)$-form $\omega$ as a sum of a $\bar\partial_{\langle A\rangle}$-harmonic term, a $\bar\partial_{\langle A\rangle}$-exact term and a $\partial_{\langle A\rangle}$-co-exact term:
\begin{equation}
\omega = \omega_\text{h} + \bar \partial_{\langle A\rangle} \alpha + \star \partial_{\langle A\rangle} \beta, \qquad \bar \partial_{\langle A\rangle} \omega_\text{h} = \partial_{\langle A\rangle} \star \omega_\text{h} = 0. \label{dgsdtsjisgf}
\end{equation}
See, e.g., \cite{Huybrechts, Voisin, Green:2012pqa} for reviews. The same holds for any $(2,2-q)$-form $\eta$, as follows from writing it as $\eta = \star \omega^\dagger$. We stress that, unless the gauge bundle is trivial, $\bar\partial_{\langle A \rangle}$-harmonic is not the same as $\partial_{\langle A \rangle}$-harmonic (where the barred and unbarred derivatives would be exchanged in \eqref{dgsdtsjisgf}). For example, $\partial_{\langle A\rangle} a_\text{h} \neq 0$ in general.
Here and in the following, the subscript ``h'' on a $(0,q)$ or $(2,2-q)$-form always refers to a $\bar\partial_{\langle A \rangle}$-harmonic form, i.e., one that satisfies \eqref{dgsdtsjisgf}.\footnote{Note, however, that the $(1,0)$-form $a_\text{h}^\dagger:= (a_\text{h})^\dagger $ is in our notation the Hermitian conjugate of $a_\text{h}$ and therefore $\partial_{\langle A \rangle}$-harmonic, not $\bar\partial_{\langle A \rangle}$-harmonic.} The two notions of harmonicity are equivalent for Hermitian forms (e.g., $\iota^* J \w \mathcal{F} = \iota^* J \w \mathcal{F}^\dagger$) and also for $(2,0)$ and $(0,2)$-forms since the latter are self-dual with respect to the Hodge star (e.g., $\Phi_\text{h}=\star \Phi_\text{h}$).

We also note that the above discussion is valid for $\langle F^{2,0}\rangle = 0$, i.e., assuming a holomorphic gauge bundle, since otherwise $\bar\partial_{\langle A \rangle}^2 \neq 0$ (except when restricting to Abelian brane fields in which case $\bar\partial_{\langle A \rangle}$ simply reduces to $\bar\partial$). We will therefore assume $\langle F^{2,0}\rangle= 0$ in the following whenever we consider non-Abelian branes. We will nevertheless keep $\langle\mathcal{F}^{2,0}\rangle$ general in the equations to track the appearance of the gauge flux in the Abelian case. Note that the 8d equations derived in the previous section are valid for fully general fluxes.

A fact we will often use below is that the Hodge decomposition is orthogonal, i.e., we can split wedge products as $\text{Tr} \int_D \eta \w \omega = \text{Tr} \int_D \left[ \eta_\text{h} \w \omega_\text{h} + \eta_\text{nh} \w \omega_\text{nh} \right]$, where ``nh'' denotes the exact and co-exact parts. Recall further that the wedge product of two harmonic forms need not be harmonic and the wedge product of two ``nh'' forms need not be ``nh'', e.g., $[\Phi_\text{h},\Phi^\dagger_\text{h}]_\text{nh}\neq 0$ and $[\Phi_\text{KK},\Phi^\dagger_\text{KK}]_\text{h}\neq 0$ in general.

It will also be convenient to adopt a more compact notation for the bulk-flux terms in the action.
Since $\Omega_{IJK}$ is covariantly constant and $\bar G_3^{3,0}$ is harmonic and therefore a constant times $\Omega$, we have that $\bar\Omega^{IJK} \bar G_{IJK}$ is constant and therefore $\bar\Omega^{klz} \bar G_{klz} |_0 = \frac{2i}{\mathcal{V}}\int_X \bar\Omega\w \bar G_3$. Here, $\mathcal{V}$ denotes the Calabi-Yau volume. We furthermore define
\begin{equation}
W_0:=\int_X \Omega\w G_3, \qquad \e^{K_\text{cs}}:=\frac{1}{4\pi|\Omega|^2\mathcal{V}}, \qquad \Xi(\Phi^\dagger):=\frac{i\sqrt{\mathcal{V}}}{8\sqrt{\pi}|\Omega|}\,\Phi^\dagger_{\bar k\bar l} \Omega^{\bar k\bar l \bar z}  G_{ij \bar z} |_0 \d z^i\w \d z^j. \label{dfuuioafjfgg}
\end{equation}
For now, the first two are just suggestive names for constants, but we will see below that they correspond to the on-shell bulk superpotential and the on-shell K\"ahler potential for the complex-structure moduli.
With this notation, we can write
\begin{equation}
\mathcal{F}^{2,0}_\phi
= \mathcal{F}^{2,0} + \frac{\sqrt{g_s}}{\sqrt{4\pi\mathcal{V}}} \left(  \Phi\, \e^{K_\text{cs}/2} \bar W_0 + \Xi(\Phi^\dagger) \right).
\end{equation}

Using these various definitions and remarks
in \eqref{faofjffegfe}, the effective action becomes\footnote{There are no terms $\sim \partial_{\langle A\rangle} a_\text{h} + \bar\partial_{\langle A\rangle} a_\text{h}^\dagger$ in the fifth line since $\iota^* J \w \partial_{\langle A\rangle} a_\text{h} = \partial_{\langle A\rangle} (  \iota^* J \w a_\text{h}) \sim \partial_{\langle A\rangle} \star a_\text{h}=0$.}
\begin{align}
S_\text{eff} &= \!\int \text{dvol}_4 \,\text{Tr} \int_D \Bigg[ - \frac{1}{2\pi} (D_\mu \Phi_\text{h}^\dagger )\w (D^\mu \Phi_\text{h})
- \frac{1}{2\pi} \star (D_\mu a_\text{h}^\dagger )\w (D^\mu a_\text{h} ) \nl \!\!
- \frac{1}{2\pi} (D_\mu \Phi_\text{KK}^\dagger )\w (D^\mu \Phi_\text{KK}) - \frac{1}{2\pi} \star (D_\mu a_\text{KK}^\dagger )\w (D^\mu a_\text{KK} ) \nl \!\!
- \frac{1}{\pi} \star \left| \left[a ,\Phi \right]_\text{h} \right|^2 
- \frac{1}{\pi} \star \left| \bar\partial_{\langle A \rangle}\Phi_\text{KK} + i\left[a ,\Phi \right]_\text{nh} \right|^2
\nl - \pi \star \left|\left( \iota^* J \w \Big( \langle\mathcal{F}^{1,1}\rangle + \frac{i}{2\pi} \{a, a^\dagger \} \Big)\right)_\text{h} + \frac{1}{2\pi}[ \Phi, \Phi^\dagger ]_\text{h} \right|^2 \nl \!\!
- \pi \star \left|\left( \iota^* J \w \Big( \langle\mathcal{F}^{1,1}\rangle + \frac{i}{2\pi} \{a, a^\dagger \} \Big) \right)_\text{nh} + \frac{1}{2\pi} \iota^* J \w \left( \partial_{\langle A \rangle} a_\text{KK} + \bar\partial_{\langle A \rangle} a_\text{KK}^\dagger  \right) + \frac{1}{2\pi}[ \Phi, \Phi^\dagger ]_\text{nh} \right|^2 \nl \!\!
- 4 \pi \star \left| \langle \mathcal{F}^{2,0}\rangle_\text{h} + \frac{i}{2\pi} \big( a^\dagger \w a^\dagger \big)_\text{h} +\frac{\sqrt{g_s}}{\sqrt{4\pi\mathcal{V}}}\! \left(  \Phi_\text{h}\, \e^{K_\text{cs}/2} \bar W_0 + \Xi_\text{h}(\Phi^\dagger) \right) \right|^2 \nl \!\!
- 4 \pi \star \left| \langle \mathcal{F}^{2,0}\rangle_\text{nh} + \frac{1}{2\pi} \partial_{\langle A \rangle} a_\text{KK}^\dagger + \frac{i}{2\pi} \big(a^\dagger \w a^\dagger\big)_\text{nh} + \frac{\sqrt{g_s}}{\sqrt{4\pi\mathcal{V}}} \!\left(  \Phi_\text{KK}\, \e^{K_\text{cs}/2} \bar W_0 + \Xi_\text{nh}(\Phi^\dagger) \right) \right|^2\! \Bigg]. \label{sdgsdgsgziui}
\end{align}

\subsection{Integrating out the KK modes}

We now claim that, in a regime of sufficiently large volumes and small brane fields, integrating out the KK modes $a_\text{KK}$ and $\Phi_\text{KK}$ amounts to discarding in \eqref{sdgsdgsgziui} all forms that are not harmonic, i.e.,
\begin{align}
S_\text{eff} &= \int \text{dvol}_4 \,\text{Tr} \int_D \Bigg[ - \frac{1}{2\pi} (D_\mu \Phi_\text{h}^\dagger )\w (D^\mu \Phi_\text{h})
- \frac{1}{2\pi} \star (D_\mu a_\text{h}^\dagger )\w (D^\mu a_\text{h} ) \nl - \frac{1}{\pi} \star \left| \left[a_\text{h} ,\Phi_\text{h} \right]_\text{h} \right|^2 - \pi \star \left| \left(\iota^* J \w \Big( \langle\mathcal{F}^{1,1}\rangle + \frac{i}{2\pi} \{a_\text{h}, a_\text{h}^\dagger \} \Big)\right)_\text{h} + \frac{1}{2\pi}[ \Phi_\text{h}, \Phi_\text{h}^\dagger ]_\text{h} \right|^2 \nl - 4 \pi \star \left| \langle \mathcal{F}^{2,0}\rangle_\text{h} + \frac{i}{2\pi}  \big( a_\text{h}^\dagger \w a_\text{h}^\dagger\big)_\text{h} + \frac{\sqrt{g_s}}{\sqrt{4\pi\mathcal{V}}} \left(\Phi_\text{h}\, \e^{K_\text{cs}/2} \bar W_0 + \Xi_\text{h}(\Phi^\dagger_\text{h})\right) \right|^2 \Bigg] \nl + \text{subleading}. \label{dgsgsgs}
\end{align}
As mentioned before, setting $\langle F^{2,0}\rangle\neq 0$ in this expression is only valid when considering Abelian brane fields.

Note that \eqref{dgsgsgs} is in general not equivalent to simply setting $a_\text{KK}=\Phi_\text{KK}=0$ in \eqref{sdgsdgsgziui}. For example, the latter would yield
$\left| \left[a_\text{h} ,\Phi_\text{h} \right]_\text{h} \right|^2 + \left| \left[a_\text{h} ,\Phi_\text{h} \right]_\text{nh} \right|^2$ in the third line of \eqref{sdgsdgsgziui}, whereas in \eqref{dgsgsgs} only $\left| \left[a_\text{h} ,\Phi_\text{h} \right]_\text{h} \right|^2$ survives. As stated before, these two expressions are not equivalent since products of harmonic forms are not necessarily harmonic. It would also not be equivalent to write $\left[a ,\Phi \right]_\text{h}$ instead of $\left[a_\text{h} ,\Phi_\text{h} \right]_\text{h}$ in \eqref{dgsgsgs} since, e.g., $\left[a_\text{KK} ,\Phi_\text{KK} \right]_\text{h}$ can be non-vanishing.

It is not immediately obvious that \eqref{dgsgsgs} is the correct result after integrating out the KK modes, in particular in fully general flux backgrounds.
On the other hand, we will see in Section \ref{4dd} that \eqref{dgsgsgs} precisely matches with the expected expressions from 4d $\mathcal{N}=1$ supergravity.
We therefore expect that there must be a regime in which \eqref{dgsgsgs} is a reliable approximation.
We will now argue that this is indeed the case. Readers who do not need to be convinced of this may proceed to Section \ref{4dd} right away.

We first consider the case where all fields are Abelian and the bulk fluxes are zero. Taking further a harmonic $\langle \mathcal{F}\rangle$, the action \eqref{sdgsdgsgziui} simplifies to
\begin{align}
S_\text{eff} &= \int \text{dvol}_4 \,\text{Tr} \int_D \Bigg[ - \frac{1}{2\pi} (D_\mu \Phi_\text{h}^\dagger )\w (D^\mu \Phi_\text{h}) 
- \frac{1}{2\pi} \star (D_\mu a_\text{h}^\dagger )\w (D^\mu a_\text{h} ) \nl - \frac{1}{2\pi} (D_\mu \Phi_\text{KK}^\dagger )\w (D^\mu \Phi_\text{KK})
- \frac{1}{2\pi} \star (D_\mu a_\text{KK}^\dagger )\w (D^\mu a_\text{KK} ) 
- \frac{1}{\pi} \star \left| \bar\partial \Phi_\text{KK} \right|^2 \nl
- \pi \star \left|\iota^* J \w \langle\mathcal{F}^{1,1}\rangle_\text{h} \right|^2
- \frac{1}{4\pi} \star \left|\iota^* J \w  \left( \partial a_\text{KK} + \bar\partial a_\text{KK}^\dagger \right) \right|^2
- 4 \pi \star \left| \langle \mathcal{F}^{2,0}\rangle_\text{h} \right|^2 \nl
- \frac{1}{\pi} \star \left| \partial a_\text{KK}^\dagger \right|^2 \Bigg]. \label{dfdalnifadlif}
\end{align}
Since the potential for the KK modes is a sum of squares, $V\propto \text{Tr} \int_D \star \big[ \left| \bar\partial \Phi_\text{KK} \right|^2 + \frac{1}{4} \big|\iota^* J \w \big( \partial a_\text{KK} + \bar\partial a_\text{KK}^\dagger \big) \big|^2 + \big| \partial a_\text{KK}^\dagger \big| ^2 + \text{const.}\big]$, it is clearly minimized at $a_\text{KK}=\Phi_\text{KK}=0$. Integrating out the KK modes is in this case therefore the same as setting them to zero in the action, with the result
\begin{align}
S_\text{eff} &= \int \text{dvol}_4 \,\text{Tr} \int_D \Bigg[ - \frac{1}{2\pi} (D_\mu \Phi_\text{h}^\dagger )\w (D^\mu \Phi_\text{h})
- \frac{1}{2\pi} \star (D_\mu a_\text{h}^\dagger )\w (D^\mu a_\text{h} ) \nl
- \pi \star \left|\iota^* J \w \langle\mathcal{F}^{1,1}\rangle_\text{h} \right|^2
- 4 \pi \star \left| \langle \mathcal{F}^{2,0}\rangle_\text{h} \right|^2 \Bigg].
\end{align}
This is consistent with \eqref{dgsgsgs} upon setting the non-Abelian and bulk-flux terms to zero there. Our claim that integrating out the KK modes yields \eqref{dgsgsgs} is therefore correct in this simple case.

Let us now consider the somewhat more complicated case where bulk fluxes are turned on (again assuming Abelian brane fields). The non-kinetic terms of \eqref{dfdalnifadlif} are then replaced by
\begin{align}
S_\text{eff} &= \int \text{dvol}_4 \,\text{Tr} \int_D \Bigg[ \ldots
- \frac{1}{\pi} \star \left| \bar\partial \Phi_\text{KK} \right|^2
- \pi \star \left|\iota^* J \w \langle\mathcal{F}^{1,1}\rangle_\text{h} \right|^2\nl
- \frac{1}{4\pi} \star \left|\iota^* J \w  \left( \partial a_\text{KK} + \bar\partial a_\text{KK}^\dagger \right) \right|^2 
- 4 \pi \star \left| \langle \mathcal{F}^{2,0}\rangle_\text{h} +\frac{\sqrt{g_s}}{\sqrt{4\pi\mathcal{V}}} \left(\Phi_\text{h}\,\e^{K_\text{cs}/2} \bar W_0 +\Xi_\text{h}(\Phi^\dagger) \right) \right|^2 \nl
- \frac{1}{\pi} \star \left| \partial a_\text{KK}^\dagger + \frac{\sqrt{\pi g_s}}{ \sqrt{\mathcal{V}}} \left(\Phi_\text{KK}\,\e^{K_\text{cs}/2} \bar W_0 + \Xi_\text{nh}(\Phi^\dagger) \right) \right|^2 
\Bigg]. \label{tuokjtujkjofj}
\end{align}
The KK modes now have a more complicated potential so that it is no longer consistent to simply set them to zero.
In particular, the part of the potential that depends on $a_\text{KK}$ is
\begin{align}
V &\propto \text{Tr}\int_D \star \left[ \frac{1}{4}\left|\iota^* J \w \left( \partial a_\text{KK} + \bar\partial a_\text{KK}^\dagger \right) \right|^2
+ \left| \partial a_\text{KK}^\dagger +\frac{\sqrt{\pi g_s}}{ \sqrt{\mathcal{V}}} \left(\Phi_\text{KK}\,\e^{K_\text{cs}/2} \bar W_0 + \Xi_\text{nh} \right) \right|^2 \right].
\end{align}
To minimize this, we make the ansatz $a_\text{KK}^\dagger = \star \bar\partial \beta$, where $\beta$ is some $(2,0)$-form. The first square then automatically vanishes since $\iota^* J \w \bar\partial a_\text{KK}^\dagger \sim \bar\partial \star a_\text{KK}^\dagger$. Without loss of generality, we can further write $\Phi_\text{KK}=\partial \gamma$, $\Xi_\text{nh}=\partial\xi$ in terms of some co-exact $(1,0)$-forms $\gamma$, $\xi$. This is because $\Phi_\text{KK}$ and $\Xi_\text{nh}$ are $(2,0)$-forms and by definition not harmonic. The potential for $a_\text{KK}$ is thus minimized by
\begin{align}
a_\text{KK}^\dagger = - \frac{\sqrt{\pi g_s}}{\sqrt{\mathcal{V}}} \left(\gamma\,\e^{K_\text{cs}/2} \bar W_0 + \xi \right).
\end{align}
Substituting the solution back into the action \eqref{tuokjtujkjofj}, we get
\begin{align}
S_\text{eff} &= \int \text{dvol}_4 \,\text{Tr} \int_D \Bigg[ \ldots
- \frac{1}{\pi} \star \left| \bar\partial \Phi_\text{KK} \right|^2
- \pi \star \left|\iota^* J \w \langle\mathcal{F}^{1,1}\rangle_\text{h} \right|^2\nl
- 4 \pi \star \left| \langle \mathcal{F}^{2,0}\rangle_\text{h} + \frac{\sqrt{g_s}}{\sqrt{4\pi \mathcal{V}}}\left(\Phi_\text{h}\,\e^{K_\text{cs}/2} \bar W_0+\Xi_\text{h}(\Phi^\dagger)\right) \right|^2
\Bigg]. \label{sdgsrrrdogsdjg}
\end{align}
One can also verify that, at large volumes, substituting the solution into the kinetic term of $a_\text{KK}$ (not displayed here) only yields subleading corrections to the kinetic terms of $\Phi_\text{h}$ and $\Phi_\text{KK}$. This is because the solution is schematically of the form $a_\text{KK} \sim \frac{\Phi_\text{h}G_3}{\sqrt{\mathcal{V}}} + \frac{\Phi_\text{KK}G_3}{\sqrt{\mathcal{V}}}$.

We still have to integrate out $\Phi_\text{KK}$ in \eqref{sdgsrrrdogsdjg}. Its potential consists of two squares (since $\Xi_\text{h}(\Phi^\dagger)$ depends on $\Phi_\text{KK}$): $V \propto \text{Tr} \int_D \star \big[ \frac{1}{\pi}\left| \bar\partial \Phi_\text{KK} \right|^2 + 4 \pi \big| \langle \mathcal{F}^{2,0}\rangle_\text{h} + \frac{\sqrt{g_s}}{\sqrt{4\pi\mathcal{V}}}\left(\Phi_\text{h}\,\e^{K_\text{cs}/2} \bar W_0+ \Xi_\text{h}\right) \big|^2 + \text{const.} \big]$.
We could try to solve this explicitly, but it suffices to check the volume scalings of the different terms and their dependence on $\Phi_\text{KK}$, $\Phi_\text{h}$. By counting the implicit metric factors in the Hodge star and in the squares, we observe that the part of the potential that depends on $\Phi_\text{KK}$ is schematically of the form
\begin{equation}
V\propto \frac{\Phi_\text{KK}^2}{\mathcal{V}^{1/3}} + \left( \mathcal{F} + \frac{\Phi_\text{h} G_3}{ \sqrt{\mathcal{V}}} + \frac{\Phi_\text{KK} G_3}{ \sqrt{\mathcal{V}}} \right)^2. \label{dgstjst}
\end{equation}
Here, we only display volume factors and brane/bulk fields and ignore the precise way the indices are contracted as well as numerical factors. To estimate the first term in the potential, we used $\int_D \star | \bar\partial \Phi_\text{KK} |^2 \sim \int_D \Delta \Phi_\text{KK} \w \Phi_\text{KK}^\dagger$ and that the eigenvalues of the Laplacian scale like $\gtrsim \frac{1}{\mathcal{V}^{1/3}}$. The scalings in \eqref{dgstjst} are reliable as long as the Calabi-Yau and the D7-brane divisor are characterized by the same length scale, i.e., for divisor volumes of the order $\mathcal{V}^{2/3}$ (as, e.g., in the T-brane-uplifting scenario of \cite{Cicoli:2015ylx}). One could also do a more refined scaling analysis distinguishing the sizes of the divisor and the Calabi-Yau, but this does not add much to our point that the KK corrections are subleading at large volumes, so we refrain from doing this here and instead continue with \eqref{dgstjst}.

It is clear from \eqref{dgstjst} that the solution will be of the form
$\Phi_\text{KK} = \frac{\mathcal{F} G_3}{\mathcal{V}^{1/6}} + \frac{\Phi_\text{h} G_3^2}{\mathcal{V}^{2/3}} + \mathcal{O}(\mathcal{V}^{-{5/6}})$.
Substituting this back into the action leads to corrections to the kinetic terms and the potential which are suppressed by at least a factor $\frac{G_3^2}{\mathcal{V}^{2/3}}$, in particular $V\propto \big(\mathcal{F} + \frac{\Phi_\text{h} G_3}{ \sqrt{\mathcal{V}}}\big)^2 \big(1 + \mathcal{O}\big(\frac{G_3^2}{\mathcal{V}^{2/3}}\big) \big)$. Hence, for large enough volumes, it is consistent to simply set $\Phi_\text{KK}=0$ in the effective action up to subleading corrections. We thus arrive at
\begin{align}
S_\text{eff} &= \int \text{dvol}_4 \,\text{Tr} \int_D \Bigg[ - \frac{1}{2\pi} (D_\mu \Phi_\text{h}^\dagger )\w (D^\mu \Phi_\text{h}) - \frac{1}{2\pi} \star (D_\mu a_\text{h}^\dagger )\w (D^\mu a_\text{h} ) \nl
- \pi \star \left|\iota^* J \w \langle\mathcal{F}^{1,1}\rangle_\text{h} \right|^2
- 4 \pi \star \left| \langle \mathcal{F}^{2,0}\rangle_\text{h} + \frac{\sqrt{g_s}}{\sqrt{4\pi\mathcal{V}}}\left(\Phi_\text{h}\,\e^{K_\text{cs}/2} \bar W_0+ \Xi_\text{h}(\Phi^\dagger_\text{h})\right) \right|^2 \Bigg]. \label{hkhjouzkgmhkh}
\end{align}
This is again consistent with our general formula \eqref{dgsgsgs}.

Last but not least, we consider the fully non-Abelian case (with $\langle F^{2,0}\rangle=0$).\footnote{Some aspects of KK modes in T-brane backgrounds were also discussed in \cite[App.~A]{Marchesano:2017kke}, but with a somewhat different focus.} Minimizing the potential of the KK modes explicitly is now much more involved than in the Abelian case due to the commutator terms. However, it will again be sufficient for our purpose to estimate the general form of the solution.
As before, we can write down the schematic form of the potential by counting the volume factors and the brane/bulk fields in \eqref{sdgsdgsgziui}.
We thus find
\begin{align}
&V\propto \frac{\big[(a_\text{h}+a_\text{KK})( \Phi_\text{h} + \Phi_\text{KK})\big]^2}{\mathcal{V}^{1/3}} + \frac{\big[\Phi_\text{KK} + (a_\text{h}+a_\text{KK})( \Phi_\text{h} + \Phi_\text{KK})\big]^2}{\mathcal{V}^{1/3}} \nll
+ \!\left(\mathcal{F}  + (a_\text{h}+ a_\text{KK})^2 + \frac{(\Phi_\text{h}+ \Phi_\text{KK})^2}{\mathcal{V}^{1/3}} \right)^{\!2}
+ \!\left(\mathcal{F} + a_\text{KK} + (a_\text{h}+ a_\text{KK})^2 + \frac{(\Phi_\text{h}+ \Phi_\text{KK})^2}{\mathcal{V}^{1/3}} \right)^{\!2} \nll
+ \!\left(\mathcal{F} + (a_\text{h}+ a_\text{KK})^2 + \frac{(\Phi_\text{h}+\Phi_\text{KK})G_3}{\sqrt{\mathcal{V}}} \right)^{\!2} +\! \left(\mathcal{F} + a_\text{KK} + (a_\text{h}+ a_\text{KK})^2 + \frac{(\Phi_\text{h}+\Phi_\text{KK})G_3}{\sqrt{\mathcal{V}}} \right)^{\!2}\!. \label{zjkozokzohoh}
\end{align}
Recall that our expansion scheme is such that we compute the effective action up to the order $\alpha'^2$,
where each factor of a brane field and each derivative counts as $\sqrt{\alpha'}$.
In this regime of small brane fields, the leading potential is
\begin{equation}
V\propto a_\text{KK}^2 + \frac{\Phi_\text{KK}^2}{\mathcal{V}^{1/3}} + \mathcal{O}(\alpha'^{3/2}).
\end{equation}
Crucially, the KK modes have vanishing vevs at this order and only get corrections which are at least quadratic in fields or derivatives. Schematically,
\begin{equation}
a_\text{KK} = 0 + \underbrace{\mathcal{F} + a_\text{h}^2 + \frac{\Phi_\text{h}^2}{\mathcal{V}^{1/3}} + \frac{\Phi_\text{h} G_3}{\sqrt{\mathcal{V}}}}_{\mathcal{O}(\alpha')} + \ldots \label{gokhgighljihligg}
\end{equation}
and similarly for $\Phi_\text{KK}$. This means that we can discard many of the couplings in \eqref{zjkozokzohoh} since they are of the order $\alpha'^{5/2}$ or higher. For example, the term $a_\text{KK}^2\Phi_\text{h}^2$ in the first line of \eqref{zjkozokzohoh} is of the order $\alpha'^3$, the term $a_\text{KK}^2\Phi_\text{KK}^2$ is of the order $\alpha'^4$, etc.
Keeping only the $\alpha'^2$ terms, the potential simplifies to
\begin{align}
V&\propto \frac{\left(a_\text{h} \Phi_\text{h}\right)^2}{\mathcal{V}^{1/3}} + \frac{\left(\Phi_\text{KK} + a_\text{h} \Phi_\text{h}\right)^2}{\mathcal{V}^{1/3}}
+ \left(\mathcal{F}  + a_\text{h}^2 + \frac{\Phi_\text{h}^2}{\mathcal{V}^{1/3}}\right)^2
+ \left(\mathcal{F} + a_\text{KK} + a_\text{h}^2 + \frac{\Phi_\text{h}^2}{\mathcal{V}^{1/3}} \right)^2 \nl
+ \left(\mathcal{F} + a_\text{h}^2 + \frac{\Phi_\text{h} G_3}{\sqrt{\mathcal{V}}} \right)^2 + \left(\mathcal{F} + a_\text{KK} + a_\text{h}^2 + \frac{\Phi_\text{h} G_3}{\sqrt{\mathcal{V}}} \right)^2 + \text{subleading}.
\end{align}
This approximation is valid for sufficiently large volumes and small brane fields.
In this regime, the leading potential for the KK modes is simply a sum of three squares (plus three terms that do not depend on the KK modes). This is easily minimized by setting each square to zero, analogously to our earlier discussion in the Abelian case. We thus arrive at
\begin{align}
V&\propto \frac{\left(a_\text{h} \Phi_\text{h}\right)^2}{\mathcal{V}^{1/3}} 
+ \left(\mathcal{F}  + a_\text{h}^2 + \frac{\Phi_\text{h}^2}{\mathcal{V}^{1/3}}\right)^2
+ \left(\mathcal{F} + a_\text{h}^2 + \frac{\Phi_\text{h} G_3}{\sqrt{\mathcal{V}}} \right)^2,
\end{align}
in agreement with our general formula \eqref{dgsgsgs}.

For later reference, we also state the solution for the KK modes. Setting to zero the corresponding squares in \eqref{sdgsdgsgziui} as explained, one finds, at leading order,
\begin{align}
&\bar\partial_{\langle A \rangle}\Phi_\text{KK} = - i \left[a_\text{h},\Phi_\text{h} \right]_\text{nh}, \label{kkmodes1} \\
&\iota^* J\w \left(\partial_{\langle A\rangle} a_\text{KK}+\bar\partial_{\langle A\rangle} a_\text{KK}^\dagger \right) = - \left( \iota^* J \w \left( 2\pi \langle\mathcal{F}^{1,1}\rangle + i \{a_\text{h} , a^\dagger_\text{h} \}\right) \right)_\text{nh} -[ \Phi_\text{h}, \Phi^\dagger_\text{h} ]_\text{nh}, \label{kkmodes2} \\
&\partial_{\langle A\rangle} a_\text{KK}^\dagger = - 2\pi \langle \mathcal{F}^{2,0}\rangle_\text{nh} -i \big( a_\text{h}^\dagger \w a_\text{h}^\dagger \big)_\text{nh}
-2\pi \frac{\sqrt{g_s}}{\sqrt{4\pi\mathcal{V}}}  \Xi_\text{nh}(\Phi_\text{h}^\dagger). \label{kkmodes3}
\end{align}

\subsection{Control over KK corrections}
\label{control}

Since we usually want to avoid computing the various KK corrections to the 4d effective action explicitly, it is desirable to have some simple 4d diagnostics for when they are under control. We now discuss this both for the corrections coming from KK couplings to bulk fluxes and for those coming from commutator terms.

We already argued above that couplings of KK modes to bulk fluxes generically lead to corrections which are suppressed if $\frac{G_3^2}{\mathcal{V}^{2/3}}\ll 1$ (as before, $G_3^2$ here refers to two powers of $G_3$ without any further implicit volume dependence).
A slightly refined version of this is to impose $\frac{g_s \e^{K_\text{cs}} |W_0|^2}{\mathcal{V}^{2/3}} \ll 1 $ and $\frac{g_s \text{Tr}|\Xi(\Phi^\dagger_\text{KK})|^2}{\mathcal{V}} \ll \frac{\text{Tr}|\Phi_\text{KK}|^2}{\mathcal{V}^{1/3}}$. One can verify by including $g_s$ in our above scaling arguments that these conditions are indeed parametrically correct so that we expect them to be useful proxies for control.
More specifically, the second condition ensures, e.g., that corrections to the potential from integrating out $\Phi_\text{KK}$ are small, as we discussed above \eqref{hkhjouzkgmhkh}. A motivation for the first condition is, e.g., that violating it would lead to mass terms of the order $\gtrsim \frac{\Phi_\text{h}^2}{\mathcal{V}^{1/3}}$ for the zero modes (cf.~\eqref{hkhjouzkgmhkh}), and so it would no longer make sense to keep the zero modes dynamical but integrate out the KK modes.

According to \eqref{dfuuioafjfgg}, we have $\text{Tr}|\Xi|^2=
\frac{\mathcal{V}}{8\pi}\text{Tr}|\Phi|^2 G_{ij \bar z} \bar G^{ij \bar z} |_0 $.
Assuming that the flux density on the divisor is of the same order as its average on $X$, we can estimate $\text{Tr}|\Xi|^2 \sim \text{Tr}|\Phi|^2 \int_X\star_6 | G_3^{2,1}|^2$. Using further $\e^{K_\text{cs}}|W_0|^2 = \frac{1}{4\pi}\int_X\star_6 |G_3^{0,3}|^2$ and that $|Q_3^\text{bulk}|\sim \int_X\star_6 |G_3|^2$ is the D3 tadpole generated by the bulk fluxes, we can write the conditions for control as
\begin{equation}
\frac{g_s \e^{K_\text{cs}} |W_0|^2}{\mathcal{V}^{2/3}} \ll 1, \qquad \frac{g_s |Q_3^\text{bulk}|}{\mathcal{V}^{2/3}} \ll 1. \label{sdgsdijgsdjlg}
\end{equation}
The first condition agrees with previous estimates for KK corrections to the bulk potential \cite{Cicoli:2013swa}. Interestingly, the second condition resembles instead the condition for controlling warping corrections (aside from the different dilaton factor), see, e.g., \cite{Junghans:2026hfm} for a discussion of the latter.\footnote{It is also interesting that the conditions for small KK corrections in 4d are stronger by a factor $\sqrt{g_s}\mathcal{V}^{1/3}$ than the corresponding conditions for small $\alpha'$ corrections in 10d. The latter are controlled if the energy density of $g_sG_3$ is small in the string frame, which translates to $\sqrt{g_s}|G_3|^2\ll 1$ in the Einstein frame or, in 4d language, $\frac{\sqrt{g_s}\e^{K_\text{cs}} |W_0|^2}{\mathcal{V}}\ll 1$, $\frac{\sqrt{g_s} Q_3^\text{bulk}}{\mathcal{V}}\ll 1$.}

In the non-Abelian case, we saw that the commutator terms in the action can lead to further KK corrections. Let us set, during the following discussion, $\langle \mathcal{F}^{2,0}\rangle$ and the bulk fluxes to zero for simplicity. We then propose that necessary conditions to generically control the KK corrections are
\begin{align}
&\int_D\star \left|\left[a_\text{h} ,\Phi_\text{h} \right] \right|^2 \ll 1, \label{kk1} \\
&\int_D\star \left|\iota^* J \w \left( 2\pi\langle\mathcal{F}^{1,1}\rangle + i \{a_\text{h}, a_\text{h}^\dagger \} \right) + [ \Phi_\text{h}, \Phi_\text{h}^\dagger ] \right|^2 \ll 1, \label{kk2} \\
&\int_D\star \left| a_\text{h}^\dagger \w a_\text{h}^\dagger  \right|^2 \ll 1. \label{kk3}
\end{align}
The reason to impose these conditions is that violating them implies that too large field values for the KK modes are sourced via \eqref{kkmodes1}--\eqref{kkmodes3}. If the first condition is violated, a field of the order $(\Phi_\text{KK})_{ij}\gtrsim \mathcal{V}^{1/6}$ is generated, and if the second or third condition is violated, we have $(a_\text{KK})_{\bar\imath}\gtrsim 1$.
For such large KK fields, it is generically no longer consistent to assume that the commutators in which they appear are subleading in the action. Consider, e.g., a violation of \eqref{kk2}. Any displacement of $\Phi_\text{h}$ with $[\Phi_\text{h},\Phi_\text{h}^\dagger]\neq 0$ that makes the left-hand side of \eqref{kk2} non-vanishing sources, because of \eqref{kkmodes2}, a KK fluctuation $a_\text{KK}=\bar\partial_{\langle A\rangle}\alpha \neq 0$. This in turn then sources a potential for $\Phi_\text{KK}$ in the third line of \eqref{sdgsdgsgziui}, leading to a KK fluctuation of the order $\Phi_\text{KK}\gtrsim [ \alpha, \Phi_\text{h} ]$. If \eqref{kk2} is violated, we have $\alpha \sim a_\text{KK} \gtrsim 1$ and therefore $\Phi_\text{KK} \gtrsim \Phi_\text{h} $, in contradiction to our initial assumption that the KK modes are small in the action compared to the zero modes.

An interesting consequence of this is that \eqref{kk2} suggests that, for backgrounds with $\iota^* J \w \langle\mathcal{F}^{1,1}\rangle\neq 0$, points near $a_\text{h}=\Phi_\text{h}=0$ in the moduli space are not controlled in the 4d EFT. This follows because $\int_D\star \left|\iota^* J \w \langle\mathcal{F}^{1,1}\rangle \right|^2 \gtrsim 1$ due to flux quantization. From the 4d perspective, this is somewhat mysterious since nothing seems to immediately go wrong near $a_\text{h}=\Phi_\text{h}=0$. However, from the 8d point of view, the vacuum is a T-brane solution with $\Phi_\text{h}\neq 0$, and the equations of motion generate an exact term in $\langle\mathcal{F}^{1,1}\rangle$ which is of the same order as the harmonic term, $\langle\mathcal{F}^{1,1}\rangle_\text{h}\sim \langle\mathcal{F}^{1,1}\rangle_\text{nh}$ \cite{Marchesano:2017kke}. In the T-brane vacuum, this exact term is precisely cancelled by $\frac{1}{2\pi}[\Phi_\text{h},\Phi_\text{h}^\dagger]_\text{nh}$, but as we move in field space away from the vacuum towards $\Phi_\text{h}\to 0$, KK modes are sourced and thus corrections to the 4d effective action become more and more important.

To arrive at the above conditions, we assumed for simplicity that the divisor volume is of the order $\mathcal{V}^{2/3}$ as before. We also assumed that harmonic/non-harmonic terms like $\left[a_\text{h} ,\Phi_\text{h} \right]_\text{h} \sim \left[a_\text{h} ,\Phi_\text{h} \right]_\text{nh}$, $\langle\mathcal{F}^{1,1}\rangle_\text{h}\sim \langle\mathcal{F}^{1,1}\rangle_\text{nh}$, etc.~are parametrically the same, which we expect to be true generically. Indeed, our arguments mainly rely on the scalings of the different terms with the volume and the brane fields, which are the same for the harmonic and non-harmonic parts. We therefore expect that \eqref{kk1}--\eqref{kk3} are reasonable diagnostics for control.
It is nevertheless possible that the conditions need to be refined in specific situations, e.g., when some of the involved commutators vanish or more parameters than just the overall Calabi-Yau volume are important for the scalings of the different terms. We leave a more careful analysis of such questions for future work. We also stress that our arguments were parametric, i.e., the ``1'' on the right-hand sides of \eqref{kk1}--\eqref{kk3} (and also \eqref{sdgsdijgsdjlg}) should be read as $\mathcal{O}(1)$. The point we wanted to make here is a qualitative one, namely that ignoring corrections from KK modes in the 4d effective action \eqref{dgsgsgs} does not only impose a lower bound on $\mathcal{V}$ via \eqref{sdgsdijgsdjlg} but can also restrict the field ranges of the brane fields.

\section{4d $D$-term and $F$-term potential}
\label{4dd}

As shown in the previous section, the effective action for the zero modes $\Phi_\text{h}$, $a_\text{h}$ is
\begin{align}
S_\text{eff} &= \int \text{dvol}_4 \,\text{Tr} \int_D \Bigg[ - \frac{1}{2\pi} (D_\mu \Phi_\text{h}^\dagger )\w (D^\mu \Phi_\text{h})
- \frac{1}{2\pi} \star (D_\mu a_\text{h}^\dagger )\w (D^\mu a_\text{h} ) \nl - \frac{1}{\pi} \star \left| \left[a_\text{h} ,\Phi_\text{h} \right]_\text{h} \right|^2 - \pi \star \left| \left( \iota^* J \w \Big( \langle\mathcal{F}^{1,1}\rangle + \frac{i}{2\pi} \{a_\text{h}, a_\text{h}^\dagger \} \Big)\right)_\text{h} + \frac{1}{2\pi}[ \Phi_\text{h}, \Phi_\text{h}^\dagger ]_\text{h} \right|^2 \nl - 4 \pi \star \left| \langle \mathcal{F}^{2,0}\rangle_\text{h} + \frac{i}{2\pi}  \big( a_\text{h}^\dagger \w a_\text{h}^\dagger\big)_\text{h} +\Phi_\text{h} \frac{\sqrt{g_s}}{ \sqrt{4\pi\mathcal{V}}} \e^{K_\text{cs}/2} \bar W_0 \right. \nl + \left. i \frac{\sqrt{g_s}}{16\pi |\Omega|} \left((\Phi_\text{h}^\dagger)_{\bar k\bar l} \Omega^{\bar k\bar l \bar z}  G_{ij \bar z} |_0 \d z^i\w \d z^j\right)_\text{h} \right|^2 \Bigg]
\end{align}
up to corrections that are subleading at sufficiently large volumes and small brane fluctuations. The final step is now to convert this to the 4d Einstein frame. Dimensionally reducing the type IIB bulk action yields the 4d Einstein-Hilbert term $2\pi \int \text{dvol}_4\, \mathcal{R}_{(4)} \mathcal{V}$, where $\mathcal{R}_{(4)}$ is the scalar curvature of $g_{\mu\nu}$. Going to the 4d Einstein frame and 4d Planck units amounts to rescaling the metric such that $2\pi \int \text{dvol}_4\, \mathcal{R}_{(4)} \mathcal{V} = \int \text{dvol}_4\,\frac{\mathcal{R}^\text{E}_{(4)}}{2}$.
This is achieved by the field redefinition $g_{\mu\nu} = \frac{g^\text{E}_{\mu\nu}}{4\pi\mathcal{V}}$, where we drop the ``E'' in the following to not clutter the expressions. We further choose to make another field redefinition $\Phi_\text{h} = \sqrt{8\pi^2\mathcal{V}}\, \hat\Phi$, $a_\text{h} = \hat a$.
This does not have a deeper meaning but is simply a convenient normalization chosen such that the kinetic terms of $\hat\Phi$ and $\hat a$ scale with the Calabi-Yau volume as in \cite{Jockers:2004yj}.

We thus obtain, in 4d Planck units,
\begin{align}
S_\text{eff} &= \int \text{dvol}_4 \,\text{Tr} \int_D \Bigg[ - (D_\mu \hat\Phi^\dagger )\w (D^\mu \hat\Phi)
- \frac{1}{8\pi^2 \mathcal{V}} \star (D_\mu \hat a^\dagger )\w (D^\mu \hat a ) \nl - \frac{1}{2\pi \mathcal{V}} \star \left| \big[\hat a ,\hat\Phi \big]_\text{h} \right|^2 - \pi \star \left| \frac{1}{4\pi\mathcal{V}} \left( \iota^* J \w \Big(  \langle\mathcal{F}^{1,1}\rangle + \frac{i}{2\pi} \{\hat a, \hat a^\dagger \} \Big)\right)_\text{h} + [ \hat\Phi , \hat\Phi^\dagger ]_\text{h} \right|^2 \nl - \frac{1}{4\pi\mathcal{V}^2} \star \left| \langle \mathcal{F}^{2,0}\rangle_\text{h} + \frac{i}{2\pi}  \big( \hat a^\dagger \w \hat a^\dagger\big)_\text{h} +\hat\Phi \sqrt{2\pi g_s}\, \e^{K_\text{cs}/2} \bar W_0 \right. \nl + \left. i \sqrt{\mathcal{V}} \frac{\sqrt{g_s}}{4\sqrt{2} |\Omega|} \left((\hat\Phi^\dagger)_{\bar k\bar l} \Omega^{\bar k\bar l \bar z}  G_{ij \bar z} |_0 \d z^i\w \d z^j\right)_\text{h} \right|^2 \Bigg]. \label{fhofdvnlolffhu}
\end{align}
We can now finally identify the non-kinetic terms with the 4d $D$-term and $F$-term potential:
\begin{align}
V_D &= \pi\, \text{Tr} \int_D \star \left| \frac{1}{4\pi\mathcal{V}} \left(\iota^* J \w \Big( \langle\mathcal{F}^{1,1}\rangle + \frac{i}{2\pi} \{\hat a, \hat a^\dagger \} \Big)\right)_\text{h} + [ \hat\Phi , \hat\Phi^\dagger ]_\text{h} \right|^2, \label{sdgsdjeoeorhjvnfjf0} \\
V_F &= \frac{1}{2\pi\mathcal{V}}\, \text{Tr} \int_D \star \left| \big[\hat a ,\hat\Phi \big]_\text{h} \right|^2 \nl + \frac{1}{4\pi\mathcal{V}^2}\, \text{Tr} \int_D \star \left| \langle \mathcal{F}^{2,0}\rangle_\text{h} + \frac{i}{2\pi}  \big( \hat a^\dagger \w \hat a^\dagger\big)_\text{h} + \hat\Phi \sqrt{2\pi g_s}\, \e^{K_\text{cs}/2}\bar W_0 \right. \nl + \left. i \sqrt{\mathcal{V}} \frac{\sqrt{g_s}}{4\sqrt{2} |\Omega|} \left((\hat\Phi^\dagger)_{\bar k\bar l} \Omega^{\bar k\bar l \bar z}  G_{ij \bar z} |_0 \d z^i\w \d z^j\right)_\text{h} \right|^2. \label{sdgsdjeoeorhjvnfjf}
\end{align}
We recall once more the caveat that we only derived this for $\langle F^{2,0}\rangle=0$, except in the special case where the brane fields are Abelian.

We could in principle stop here but it is useful for many applications to rewrite $V_D$ and $V_F$ such that the structure of 4d $\mathcal{N}=1$ supergravity (in terms of the K\"ahler potential, the superpotential and the gauge-kinetic functions) becomes manifest. For this last part of the paper, we will specialize to Abelian worldvolume fluxes for simplicity, i.e., we take the gauge bundle to be a direct sum of line bundles, $\mathcal{L}_1\oplus \mathcal{L}_2 \oplus \ldots \mathcal{L}_N$, so that $\langle A\rangle$, $\langle \mathcal{F}\rangle$ are valued in the Cartan subalgebra of the gauge algebra.\footnote{The orientifold projection constrains these line bundles (see \cite[Sec.~2]{Junghans:2026hfm} for a review) but this is not important for the discussion.} This is the most studied situation in the string-compactification literature (see, e.g., \cite{Blumenhagen:2008zz}) and applies in particular to T-brane backgrounds, which are the main motivation for this work (see, e.g., \cite{Cecotti:2010bp, Cicoli:2015ylx, Marchesano:2017kke}). Note that only the background fluxes will be taken Abelian from now on, whereas the dynamical fields $\hat a$, $\hat\Phi$ are kept fully non-Abelian. The previous results \eqref{sdgsdjeoeorhjvnfjf0}, \eqref{sdgsdjeoeorhjvnfjf} are valid without this assumption.

The key step is to expand the squares in $V_D$ and $V_F$ in bases of appropriate bundle-valued cohomology groups. We now do this explicitly for the second square in the $F$-term potential \eqref{sdgsdjeoeorhjvnfjf}. We can write it as
\begin{align}
V_F 
&= \ldots + \frac{g_s\,\e^{K_\text{cs}}}{2\mathcal{V}^2} \text{Tr} \int_D \Bigg[ \frac{\sqrt{2}|\Omega|\sqrt{\mathcal{V}}}{\sqrt{g_s}} \left(\langle \mathcal{F}^{2,0}\rangle_\text{h} + \frac{i}{2\pi}  \big( \hat a^\dagger \w \hat a^\dagger\big)_\text{h}\right) + \hat\Phi \bar W_0 \nl + i \mathcal{V} \left( (\hat\Phi^\dagger)_{\bar 1\bar 2} \Omega^{\bar 1\bar 2\bar z } G_{12 \bar z}|_0 \d z^1 \w \d z^2 \right)_\text{h} \Bigg] \w \Bigg[\frac{\sqrt{2}|\Omega|\sqrt{\mathcal{V}}}{\sqrt{g_s}} \left(\langle \mathcal{F}^{0,2}\rangle_\text{h} + \frac{i}{2\pi}  \big( \hat a \w \hat a\big)_\text{h}\right) \nl + \hat\Phi^\dagger W_0 - i \mathcal{V} \left({\hat\Phi}_{12} \bar\Omega^{12 z }  \bar G_{\bar 1 \bar 2 z} |_0 \d \bar z^{\bar 1} \w \d \bar z^{\bar 2} \right)_\text{h} \Bigg], \label{fajfgejgf}
\end{align}
where we used that $(2,0)$-forms are self-dual with respect to $\star$.
Since we take the gauge bundle to be a sum of line bundles, the $[\ldots ]_{ab}$ component of an adjoint $\bar\partial_{\langle A\rangle}$-harmonic $(0,q)$-form is an element of $H^q(D,\mathcal{L}_a \otimes \mathcal{L}_b^{-1})$, where $a,b=1,\ldots,N$ are the gauge indices. Accordingly, we expand each $[\ldots]_{ab}$ component of the $(0,2)$-forms in \eqref{fajfgejgf} in a basis of $H^2(D,\mathcal{L}_{a} \otimes \mathcal{L}_{b}^{-1})$. The corresponding expansion of the $(2,0)$-forms is obtained by complex conjugation. For example, $\hat\Phi$ is expanded as
\begin{equation}
\hat \Phi_{ab} = \varphi^{A} \, (\omega_{A})_{ab}, \qquad (\hat \Phi^\dagger)_{ab} = \bar \varphi^{\bar A} \, (\omega_{\bar A}^\dagger)_{ab},
\end{equation}
where the $\varphi^A$'s are the 4d moduli and the $\omega^\dagger_{\bar {A}}$'s are a basis of matrix-valued $(0,2)$-forms. We define the $\omega^\dagger_{\bar {A}}$'s such that each of them has one non-vanishing matrix entry $(\omega^\dagger_{\bar {A}})_{ab}\in H^2(D,\mathcal{L}_{a} \otimes \mathcal{L}_{b}^{-1})$ for some $a,b$ and all other matrix entries are zero. Hence, there are $\text{dim} H^{0,2}(D)$ basis forms with non-zero entry at $a=b=1$, $\text{dim} H^2(D,\mathcal{L}_{2} \otimes \mathcal{L}_{1}^{-1})$ basis forms with non-zero entry at $a=2,b=1$, and so on. The index ${\bar {A}}$ thus runs from 1 to $\sum_{a,b}\text{dim} H^2(D,\mathcal{L}_{a} \otimes \mathcal{L}_{b}^{-1})$.\footnote{Some of the $\varphi^A$'s are projected out depending on the orientifold projection, see \cite[Sec.~2]{Junghans:2026hfm} for a discussion.}

Defining $K_{A\bar B}:= \text{Tr} \int_D \omega_A \w \omega^\dagger_{\bar B}$ and denoting by $K^{\bar B A}$ the inverse, we can write any wedge product of harmonic $(2,0)$ and $(0,2)$-forms as
\begin{align}
\text{Tr}\int_D \big[ \ldots\big] \w \big[ \ldots\big] &= K^{\bar B A}\, \text{Tr} \int_D \big[ \ldots \big] \w \omega^\dagger_{\bar B}\, \text{Tr} \int_D \big[ \ldots \big] \w \omega_A. \label{tokkklghjghgjh}
\end{align}
Using this in $V_F$, we obtain
\begin{align}
V_F &= \ldots + \frac{g_s\,\e^{K_\text{cs}} K^{\bar B A}}{2\mathcal{V}^2} \text{Tr} \int_D \Bigg[ \frac{\sqrt{2}|\Omega|\sqrt{\mathcal{V}}}{\sqrt{g_s}} \left(\langle \mathcal{F}^{2,0}\rangle_\text{h} + \frac{i}{2\pi}  \big( \hat a^\dagger\w \hat a^\dagger\big)_\text{h}\right) + \hat\Phi \bar W_0 \nl + i \mathcal{V}\left( (\hat\Phi^\dagger)_{\bar 1\bar 2} \Omega^{\bar 1\bar 2\bar z } G_{12 \bar z}|_0 \d z^1 \w \d z^2 \right)_\text{h} \Bigg] \w \omega^\dagger_{\bar B}\, \text{Tr} \int_D \Bigg[\frac{\sqrt{2}|\Omega|\sqrt{\mathcal{V}}}{\sqrt{g_s}} \left(\langle \mathcal{F}^{0,2}\rangle_\text{h} + \frac{i}{2\pi}  \big( \hat a\w \hat a\big)_\text{h}\right) \nl +  \hat\Phi^\dagger W_0 - i \mathcal{V}\left( {\hat\Phi}_{12} \bar\Omega^{12 z }  \bar G_{\bar 1 \bar 2 z} |_0 \d \bar z^{\bar 1} \w \d \bar z^{\bar 2}\right)_\text{h} \Bigg] \w \omega_{A}. \label{dsgsdgsdgsgsgd}
\end{align}
This expression already looks suspiciously like an $F$-term potential $V_F \sim K^{A\bar B} (D_A W) (D_{\bar B}\bar W)$. We will indeed see below that it can be obtained from a K\"ahler potential and a superpotential as expected. Also the suggestive name $K_{A\bar B}$ chosen above is not a coincidence: we will see that $K_{A\bar B}$ is the K\"ahler metric of the $\varphi^A$ moduli space.

In principle, we could perform an analogous expansion of the first term in \eqref{sdgsdjeoeorhjvnfjf} in terms of harmonic $(2,1)$ and $(0,1)$-forms. We refrain from spelling this out here since \eqref{dsgsdgsdgsgsgd} is sufficient to determine the K\"ahler potential and the superpotential.

We instead move on to the $D$-term potential \eqref{sdgsdjeoeorhjvnfjf0}, which is expanded into harmonic $(0,0)$-forms and their Hodge-dual $(2,2)$-forms. To this end, we abbreviate $V_D = \pi\, \text{Tr} \int_D \star \rho^2$, where $\rho:= \star \big[ \frac{1}{4\pi\mathcal{V}}\big( \iota^* J \w \langle\mathcal{F}^{1,1}\rangle \big)_\text{h} + \ldots \big]$ is a matrix-valued harmonic $(0,0)$-form. Since $\rho$ is Hermitian, it is not only $\bar\partial_{\langle A\rangle}$-harmonic but also $\partial_{\langle A\rangle}$-harmonic. This is only possible if $[\langle F^{1,1} \rangle,\rho]=0$, as one can verify using $\{\partial_{\langle A\rangle}, \bar\partial_{\langle A\rangle}\}\rho=0$. Since $\langle F^{1,1} \rangle$ is valued in the Cartan of the gauge algebra, we can choose a basis for the generators such that $\langle F^{1,1} \rangle$ is only non-vanishing along the diagonal, i.e.,
\begin{equation}
\langle F^{1,1} \rangle = \begin{pmatrix}
                        \langle F^{1,1} \rangle_{1,1}  & 0 &\cdots & 0 \\
                        0 & \langle F^{1,1} \rangle_{2,2}  & \cdots & 0 \\
                        \vdots &  &  & \vdots \\
                        0 & 0 &\cdots & \langle F^{1,1} \rangle_{N,N}
                          \end{pmatrix} \sim \begin{pmatrix}
                        c_1(\mathcal{L}_1)  & 0 &\cdots & 0 \\
                        0 & c_1(\mathcal{L}_2) & \cdots & 0 \\
                        \vdots &   &  & \vdots \\
                        0 & 0 &\cdots  & c_1(\mathcal{L}_N)
                          \end{pmatrix},
\end{equation}
where $c_1(\mathcal{L}_a)$ is the first Chern class of the $a$th line bundle. In the case where all matrix entries are different, $\rho$ can only commute with $\langle F^{1,1} \rangle$ if it is also valued in the Cartan. It then follows that $[\langle A\rangle,\rho ]=0$ and therefore $\rho$ is not only closed with respect to $\bar\partial_{\langle A\rangle}$ and $\partial_{\langle A\rangle}$ but also with respect to $\bar\partial$ and $\partial$, i.e., $\rho$ is constant. In the case where $n$ entries of the $\langle F^{1,1} \rangle$ matrix are equal, $\rho$ is allowed to have off-diagonal entries. However, then also $n$ entries of $\langle A \rangle$ are equal (up to flat connections which we absorb into $\hat a+\hat a^\dagger$ without loss of generality) so that $\rho$ still commutes with $\langle A \rangle$ and is therefore still constant.
In conclusion,
\begin{equation}
\bar\partial\rho = \partial\rho = 0, \qquad [\langle F^{1,1}\rangle,\rho] = [\langle A\rangle,\rho] = 0.
\end{equation}
Using this, we can write
\begin{align}
V_D &= \pi\, \text{Tr} \int_D \star \rho^2 = \frac{\pi}{\int_D \star 1}\, \text{Tr} \left(\int_D \star \rho \right)^2 \nll
= \frac{2\pi}{ \int_D \iota^* J^2} \text{Tr} \left(
\frac{1}{4\pi\mathcal{V}}\, \int_D  \iota^* J \w \left( \langle\mathcal{F}^{1,1}\rangle + \frac{i}{2\pi} \{\hat a, \hat a^\dagger \} \right) + \int_D [ \hat\Phi , \hat\Phi^\dagger ]
\right)^2, \label{dsosgdsogghjhijkvc}
\end{align}
where we used $\int_D \star 1 = \frac{1}{2}\int_D \iota^* J^2$. This now has the familiar form of the 4d $D$-term potential dictated by the structure of $\mathcal{N}=1$ supergravity. In particular, the terms in the square are the 4d $D$-terms and the denominator can be identified with the gauge-kinetic function up to normalization, $\text{Re} f_\text{D7}\sim \int_D \iota^* J^2$. One can verify that \eqref{dsosgdsogghjhijkvc} has the same form as previous expressions in the literature, e.g., the potential used in \cite{Cicoli:2015ylx}.

\subsection{K\"ahler potential}

Obtaining the K\"ahler potential is straightforward: At the order in $\alpha'$ we consider for $V_F$, only quadratic terms in $\hat\Phi$ in the K\"ahler potential are relevant. The only gauge-invariant and real term one can write down is then $\text{Tr} \int_D \hat\Phi^\dagger \w \hat\Phi$, up to normalization. Analogous considerations apply to $\hat a$. We can thus take the K\"ahler potential to be
\begin{align}
K &= \text{Tr}\int_D \hat\Phi^\dagger \w \hat\Phi + \frac{1}{8\pi^2 \mathcal{V}} \text{Tr}\int_D \star \hat a^\dagger \w \hat a - 2\ln\mathcal{V} - \ln \frac{2}{g_s} - \ln \left( |\Omega|^2 \mathcal{V} \right) - \ln(4\pi). \label{dsgfsgfiljafeeee}
\end{align}
Here we added a constant term that is independent of the brane moduli. As we will see below, this is necessary to match with our above expression for $V_F$. Note further that this term agrees precisely with the known K\"ahler potential for the bulk moduli at the order $\alpha'^0$ \cite{Jockers:2004yj, Jockers:2005zy} (using $-i\int_X \Omega \w \bar\Omega=|\Omega|^2 \mathcal{V}$).\footnote{At $\mathcal{O}(\alpha')$, the on-shell K\"ahler potential for the axio-dilaton and the K\"ahler moduli differs from $- 2\ln\mathcal{V} - \ln \frac{2}{g_s}$ due to brane-moduli-dependent shifts in the K\"ahler coordinates \cite{Jockers:2004yj, Jockers:2005zy} but this is not relevant for us since an $\mathcal{O}(\alpha')$ correction to the constant in \eqref{dsgfsgfiljafeeee} corrects $V_F$ at $\mathcal{O}(\alpha'^3)$, which is beyond the order we compute.} See also \cite[App.~A]{Junghans:2026hfm} for a discussion.
This is also the reason why we chose the notation $\e^{K_\text{cs}}:=\frac{1}{4\pi|\Omega|^2 \mathcal{V}}$ earlier: $K_\text{cs}$ is simply the on-shell K\"ahler potential of the complex-structure moduli.
For later convenience, we also note that
\begin{equation}
\e^K = \frac{g_s\,\e^{K_\text{cs}}}{2 \mathcal{V}^2} + \mathcal{O}(\alpha'). \label{kjfhjklfjkfj}
\end{equation}

As we explained before, the 4d brane moduli are extracted from the matrix-valued $(2,0)$-form $\hat\Phi$ by expanding each entry of the matrix into bundle-valued harmonics, i.e., $\hat\Phi_{ab} = \varphi^A (x^\mu) (\omega_A)_{ab}$. The K\"ahler metric is thus
\begin{equation}
\frac{\partial}{\partial\varphi^A} \frac{\partial}{\partial\bar \varphi^{\bar B} } K = K_{A\bar B}, \label{dfldsghgdsghukgn}
\end{equation}
where $K_{A\bar B}$ was defined above \eqref{tokkklghjghgjh}.
Let us also check that we chose the correct normalization of $K$. To this end, we compute the kinetic term: Using \eqref{dfldsghgdsghukgn}, it is $\mathcal{L}_\text{kin} = - \left(\frac{\partial}{\partial\varphi^A} \frac{\partial}{\partial\bar \varphi^{\bar B}} K \right) \left(D_\mu \varphi^A\right)\left(D^\mu \bar \varphi^{\bar B} \right) = - K_{A\bar B} \left(D_\mu \varphi^A \right) \left(D^\mu \bar \varphi^{\bar B}\right) = -\text{Tr}\int_D D_\mu \hat\Phi^\dagger \w D^\mu \hat\Phi $. This agrees with the kinetic term in \eqref{fhofdvnlolffhu} obtained from the dimensional reduction so that our choice of normalization is indeed the correct one. Determining the K\"ahler potential for the Wilson-line moduli $\hat a$ proceeds analogously.

\subsection{Superpotential}

We now move on to the superpotential. The general $F$-term potential is of the form $V_F=\e^K (|D_IW|^2-3|W|^2)$, where the index $I$ runs over all bulk and brane moduli. The latter include the deformation moduli $\varphi^A$ we already discussed (i.e., the expansion coefficients of $\hat\Phi$) as well as the Wilson-line moduli (i.e., the coefficients of an expansion of $\hat a$ into harmonic $(0,1)$-forms, which we need not spell out here).

Assuming that the axio-dilaton and complex-structure moduli are integrated out supersymmetrically (corresponding to choosing ISD bulk fluxes), we have $D_SW=D_{Z_i} W=0$.
The index $I$ in $V_F$ thus only runs over the K\"ahler and brane moduli. Due to the well-known no-scale structure of the GKP background, we further have that the $F$-terms of the K\"ahler moduli and the Wilson-line moduli cancel the $-3\e^K|W|^2$ term in $V_F$ \cite{Giddings:2001yu, Grimm:2004uq}. More precisely, the cancelation is only perfect if the superpotential does not depend on the Wilson-line moduli but otherwise leaves behind a term $\e^K |\partial_a W|^2$.\footnote{To see that the Wilson-line moduli participate in the no-scale structure, one can compare the results of, e.g., \cite{Grana:2003ek} and \cite{Jockers:2004yj} to observe that the Wilson-line moduli mix with the K\"ahler moduli in the K\"ahler potential in exactly the same way as the D3-brane moduli, which are indeed known to participate in the no-scale structure. We thus have $K_IK^{I\bar J}K_{\bar J}=3$, where $I$ here runs over the K\"ahler moduli, the D3-brane moduli and the Wilson-line moduli. Hence, the Wilson-line moduli only contribute to $V_F$ if $\partial_a W \neq 0$. In particular, they do not contribute in the Abelian case, consistently with our dimensional reduction (see also \cite{Camara:2004jj}). One can also check explicitly, e.g., in models with one K\"ahler modulus that $K^a=0$ so that $\e^K \left( \partial_a W K^a \bar W + \text{c.c}\right)=0$ and the only remaining contribution from the K\"ahler/Wilson-line-moduli $F$-terms to $V_F$ is $\e^K |\partial_a W|^2$.} After imposing the no-scale structure, the full $F$-term potential is thus $V_F= \e^K |\partial_a W|^2 + \e^K |D_{\varphi} W|^2$. Here we denoted the Wilson-line moduli collectively by $a$ and the deformation moduli collectively by $\varphi$. See also \cite[App.~A]{Junghans:2026hfm} for a related discussion.

The two terms in $V_F$ can now be matched with the two squares in \eqref{sdgsdjeoeorhjvnfjf}. To derive the superpotential, it is actually sufficient to do this exercise for the second term.
We can thus write
\begin{equation}
V_F = \ldots + \e^K \left( D_{\varphi^{A}} W \right) \left( D_{\bar \varphi^{\bar B}} \bar W\right) K^{A\bar B} . \label{sdgsdfsdjjjj}
\end{equation}
We claim that, at the order $\alpha'^2$, this is identical to \eqref{dsgsdgsdgsgsgd} for the superpotential\footnote{One may think that the last term in $W$ is not holomorphic since it depends on $\bar \Omega$ and $\bar G_3$, which are functions of $\bar Z_i$ and $\bar S$, respectively. However, recall that we compute the superpotential for the brane moduli at \emph{fixed} bulk moduli. We will see below that our $W$ matches perfectly with previously obtained expressions for the Abelian brane superpotential.}
\begin{align}
W &= W_0 +
\frac{\sqrt{2}|\Omega|\sqrt{\mathcal{V}}}{\sqrt{g_s}} \text{Tr} \int_D \left(\langle \mathcal{F}^{0,2}\rangle_\text{h} + \frac{i}{2\pi}  \big( \hat a \w \hat a\big)_\text{h}\right) \w \hat\Phi \nl - \frac{i}{2} \mathcal{V}\, \text{Tr} \int_D \hat\Phi \w \Big(\hat\Phi_{12} \bar\Omega^{12z} \bar G_{\bar 1\bar 2 z} |_0 \d \bar z^{\bar 1} \w \d \bar z^{\bar 2}\Big)_\text{h} + \mathcal{O}(\alpha'^2). \label{dafraelirjiar}
\end{align}
Indeed, the $F$-term resulting from \eqref{dsgfsgfiljafeeee}, \eqref{dafraelirjiar} is
\begin{align}
\frac{\partial}{\partial\varphi^A} W + \left(\frac{\partial}{\partial\varphi^A} K\right) W &
= \text{Tr} \int_D \omega_A \w \Bigg( \frac{\sqrt{2}|\Omega|\sqrt{\mathcal{V}}}{\sqrt{g_s}} \left(\langle \mathcal{F}^{0,2}\rangle_\text{h} + \frac{i}{2\pi}  \big( \hat a\w \hat a\big)_\text{h}\right) \nl  - i \mathcal{V} \Big({\hat\Phi}_{12} \bar \Omega^{12 z }\bar G_{\bar 1\bar 2 z } |_0 \d \bar z^{\bar 1}\w \d \bar z^{\bar 2}\Big)_\text{h} + \hat\Phi^\dagger W_0 \Bigg) + \mathcal{O}(\alpha'^{3/2}). \label{dsgggosdgjsgsnlig}
\end{align}
Using this together with \eqref{kjfhjklfjkfj} in \eqref{sdgsdfsdjjjj}, we recover precisely \eqref{dsgsdgsdgsgsgd}. Note that, since we computed $V_F$ at the order $\alpha'^2$, this fixes the $\hat\Phi$-dependent part of $W$ until the order $\alpha'^{3/2}$ and the constant part of $W$ until the order $\sqrt{\alpha'}$. Higher-order corrections to $W$ would enter $V_F$ earliest at the order $\alpha'^3$ and are therefore not fixed by our analysis.
Also note that we can drop the ``h'' subscripts in \eqref{dafraelirjiar} since the non-harmonic parts of the forms integrate to zero anyway.

According to \eqref{dafraelirjiar}, a $\hat\Phi$ dependence in $W$ is generated, at the order to which we compute it, by several effects: $(0,2)$-form worldvolume flux\footnote{As stated before, our derivation strictly speaking assumed $\langle F^{0,2}\rangle=0$ except in the case of Abelian branes.}, non-Abelian Wilson lines and $(2,1)$-form bulk fluxes. The latter enter the superpotential because the fluctuating worldvolume flux $\mathcal{F}_\phi=\iota_\phi^* B_2 + \frac{1}{2\pi}F_2$ appearing in the brane action contains the pull-back of the $B_2$ field. In a background with bulk fluxes, the $B_2$ field varies as we deform the branes and thus $\mathcal{F}_\phi$ can be non-zero at finite $\hat\Phi$ even when it vanishes at $\hat\Phi=0$. In the Abelian case, this effect was already observed in \cite{Gomis:2005wc} (see also \cite{Martucci:2006ij}).

Let us also check that our superpotential \eqref{dafraelirjiar} reduces to the known Abelian expression for the case of a single D7-brane.
The Abelian flux superpotential including bulk and brane contributions is \cite{Gukov:1999ya, Jockers:2005zy, Martucci:2006ij, Denef:2008wq, Arends:2014qca}
\begin{align}
W &= \int_X \Omega \w G_3 + \int_\Gamma \tilde{\mathcal{F}} \w \Omega. \label{sgsdligsdg}
\end{align}
Here, $\Gamma$ denotes a 5-chain stretching between the undeformed brane divisor at $z=0$ and the deformed one at $z=\frac{\phi^z}{2\pi}$.
Furthermore, $\tilde{\mathcal{F}}$ denotes the extension of the worldvolume flux to $\Gamma$.
The brane-moduli dependence of $W$ is thus implicit in its second term.
Note that, in \eqref{sgsdligsdg}, we ignore a backreaction term found in \cite{Denef:2008wq, Arends:2014qca} which we expect to be subleading in $V_F$ at small $g_s$.

For small deformations of the brane divisor, we can explicitly evaluate $W$ by expanding $\tilde{\mathcal{F}}^{0,2} = \left(\mathcal{F}_{\bar 1\bar 2} + z H_{\bar 1 \bar 2 z}|_0 + \bar z H_{\bar 1 \bar 2 \bar z}|_0 + \ldots \right) \d \bar z^{\bar 1} \w \d  \bar z^{\bar 2}$ and $\Omega = (\Omega_{12 z}|_0 + \partial_z \Omega_{12z}|_0 z + \ldots) \d z \w \d z^1 \w \d z^2$. The 5-chain $\Gamma$ is given by $D$ times an interval $t\in [0,1]$, where we parametrize $z(t)=t \phi^z/2\pi$. Hence,
\begin{align}
\int_\Gamma \tilde{\mathcal{F}} \w \Omega &= \frac{\phi^z}{2\pi} \int_0^1 \d t \int_D \left( \mathcal{F}_{\bar 1 \bar 2} + z(t) H_{\bar 1 \bar 2 z}|_0 + \bar z(t) H_{\bar 1 \bar 2 \bar z}|_0 + \ldots \right) \left( \Omega_{12z}|_0 + \ldots \right)  \nl \d z^1 \w \d z^2 \w \d \bar z^{\bar 1} \w \d  \bar z^{\bar 2} \nll
= \int_D \left( \mathcal{F}_{\bar 1 \bar 2} \Omega_{12z}|_0 \frac{\phi^z}{2\pi}
- \frac{ig_s}{2} \bar G_{\bar 1 \bar 2 z} \Omega_{12z}|_0 \frac{(\phi^z)^2}{8\pi^2}
+ H_{\bar 1 \bar 2 \bar z} \Omega_{12z}|_0 \frac{\phi^z \phi^{\bar z}}{8\pi^2}
\right)\nl \d z^1 \w \d z^2 \w \d \bar z^{\bar 1} \w \d  \bar z^{\bar 2} +\mathcal{O}(\alpha'^2) \nll
= \frac{|\Omega|}{2\pi\sqrt{g_s}} \int_D \mathcal{F}^{0,2} \w \Phi
 - \frac{i}{16\pi^2} \int_D \Phi \w \Phi_{12} \bar \Omega^{12 z} \bar G_{\bar 1 \bar 2 z}|_0 \d \bar z^{\bar 1} \w \d  \bar z^{\bar 2} \nl
- \frac{i}{8\pi^2 \mathcal{V} g_s} \int_X \Omega \w H_3 \int_D \Phi^\dagger \w \Phi + \mathcal{O}(\alpha'^2).
\label{dsgslidf}
\end{align}
We now discard any non-harmonic pieces in $\Phi$ since, as discussed in Section \ref{harmo}, they would correspond to higher-order corrections in the regime we consider.
We can thus set, at the leading order, $\Phi = \Phi_\text{h} = \sqrt{8\pi^2\mathcal{V}}\,\hat\Phi$. This implies that we can also replace $\mathcal{F}^{0,2}$ and $\Phi_{12} \bar \Omega^{12 z} \bar G_{\bar 1 \bar 2 z}|_0 \d \bar z^{\bar 1} \w \d  \bar z^{\bar 2}$ by their harmonic parts under the integral.
We thus obtain
\begin{align}
\int_\Gamma \tilde{\mathcal{F}} \w \Omega &= \frac{\sqrt{2}|\Omega|\sqrt{\mathcal{V}}}{\sqrt{g_s}} \int_D \langle \mathcal{F}^{0,2}\rangle_\text{h} \w \hat\Phi
 - \frac{i}{2} \mathcal{V} \int_D \hat\Phi \w \left(\hat\Phi_{12} \bar \Omega^{12 z} \bar G_{\bar 1 \bar 2 z}|_0 \d \bar z^{\bar 1} \w \d  \bar z^{\bar 2}\right)_\text{h} \nl - \frac{i}{g_s} \int_X \Omega \w H_3 \int_D \hat\Phi^\dagger \w \hat\Phi + \mathcal{O}(\alpha'^2). \label{dsgslidf}
\end{align}
Substituting this into \eqref{sgsdligsdg}, we get
\begin{align}
W &= \int_X \Omega \w \left(F_3 - SH_3\right) \nl + \frac{\sqrt{2}|\Omega|\sqrt{\mathcal{V}}}{\sqrt{g_s}} \int_D \langle \mathcal{F}^{0,2}\rangle_\text{h} \w \hat\Phi
 - \frac{i}{2} \mathcal{V} \int_D \hat\Phi \w \left( \hat\Phi_{12} \bar \Omega^{12 z} \bar G_{\bar 1 \bar 2 z}|_0 \d \bar z^{\bar 1} \w \d  \bar z^{\bar 2} \right)_\text{h} + \mathcal{O}(\alpha'^2). \label{gdslijsdjilghligs}
\end{align}
Here we introduced the shifted K\"ahler coordinate $S=\tau + \frac{i}{g_s} \int_D \hat\Phi^\dagger\w\hat\Phi + \mathcal{O}(\alpha'^2)$ \cite{Jockers:2004yj, Jockers:2005zy}, where $\tau= C_0+ \frac{i}{g_s}$ is the axio-dilaton and $G_3 =F_3-\tau H_3$ in terms of the RR and NSNS field strengths $F_3$ and $H_3$. Note that, without absorbing $\int_D \hat\Phi^\dagger\w\hat\Phi$ into the definition of $S$, one would erroneously conclude that $W$ is non-holomorphic in the brane moduli, which would be inconsistent.

Now that we have made the $\hat\Phi$ dependence of \eqref{sgsdligsdg} explicit, we can verify that it matches with our result for $W$ from the dimensional reduction.
Since the first line in \eqref{gdslijsdjilghligs} only depends on the bulk moduli\footnote{This is true off-shell. After substituting the solutions of $D_SW=D_{Z_i}W=0$ for $S$, $Z_i$, the term can depend on the brane moduli.}, it does not enter $V_F$ through $\frac{\partial }{\partial\varphi^A} W$ but only through $\frac{\partial K}{\partial\varphi^A} W$. As we mentioned earlier, such $\varphi^A$-independent terms in $W$ are fixed by our analysis until the order $\sqrt{\alpha'}$. In particular, the shift in $S$ arises in \eqref{gdslijsdjilghligs} at the order $\alpha'^{3/2}$ and can therefore be discarded.
To compare \eqref{gdslijsdjilghligs} with our result, we thus have to replace $\int_X \Omega \w \left(F_3 - SH_3\right)$ by $W_0=\int_X \Omega \w G_3$. Doing this, we find that \eqref{gdslijsdjilghligs} perfectly matches with \eqref{dafraelirjiar} in the Abelian case. Moreover, by analogy with the Abelian case, we can more compactly write our non-Abelian $W$ as
\begin{align}
W &= \int_X \Omega \w G_3 + \text{Tr} \int_\Gamma \tilde{\mathcal{F}} \w \Omega + \mathcal{O}(\alpha'^2). \label{dgiadgiadgf}
\end{align}
Hence, the non-Abelian superpotential is simply the Abelian one with a gauge trace in front. This result is consistent with previous proposals for non-Abelian extensions of $W$ in \cite{Marchesano:2009rz, Marchesano:2010bs}, which involve the symmetrized trace STr. At the order we considered in our calculation, the symmetrized trace is equivalent to the ordinary one. In fact, \eqref{dgiadgiadgf} is the only gauge-invariant expression at this order that is consistent with the Abelian result \eqref{sgsdligsdg} (since, e.g., the trace of a commutator vanishes). At higher orders in the brane fields, there are a priori multiple possibilities for a non-Abelian $W$ and it would be interesting to check which is the right one.

\section{Conclusion}
\label{concl}

In this note, we computed the $D$-term and $F$-term potential of a stack of D7-branes wrapped on a divisor of a Calabi-Yau orientifold, at the order $\alpha'^2$ (where each brane field and each derivative counts as $\sqrt{\alpha'}$) and in the probe approximation (i.e., ignoring the backreaction of localized D7/O7 and D3/O3 sources onto the bulk, aside from imposing the tadpole conditions). We derived the full potential in one go and from first principles, by dimensionally reducing the non-Abelian DBI action of the D7-branes.
The result is
\begin{subequations}
\setlength{\fboxsep}{1em}
\begin{empheq}[box=\fbox]{align}
V_D &= \frac{2\pi}{ \int_D \iota^* J^2} \text{Tr} \left(
\frac{1}{4\pi\mathcal{V}}\, \int_D  \iota^* J \w \left( \langle\mathcal{F}^{1,1}\rangle + \frac{i}{2\pi} \{\hat a, \hat a^\dagger \} \right) + \int_D [ \hat\Phi , \hat\Phi^\dagger ]
\right)^2 + \mathcal{O}(\alpha'^3), \notag \\
K &= \text{Tr}\int_D \hat\Phi^\dagger \w \hat\Phi + \frac{1}{8\pi^2 \mathcal{V}} \text{Tr}\int_D \star \hat a^\dagger \w \hat a - 2\ln\mathcal{V} - \ln \frac{2}{g_s} - \ln \left( |\Omega|^2 \mathcal{V} \right) - \ln(4\pi) + \mathcal{O}(\alpha'^2), \notag \\
W &= \int_X \Omega \w G_3 + \frac{\sqrt{2}|\Omega|\sqrt{\mathcal{V}}}{\sqrt{g_s}} \text{Tr} \int_D \left(\langle\mathcal{F}^{0,2}\rangle + \frac{i}{2\pi}\hat a\w \hat a \right) \w \hat\Phi \nl - \frac{i}{2} \mathcal{V}\, \text{Tr} \int_D \hat\Phi \w \hat\Phi_{12} \bar\Omega^{12z} \bar G_{\bar 1\bar 2 z} |_0 \d \bar z^{\bar 1} \w \d \bar z^{\bar 2} + \mathcal{O}(\alpha'^2), \notag 
\end{empheq}
\end{subequations}
where the worldvolume fluxes (but not the brane fields) are taken Abelian and we refer to the main text for the precise assumptions and definitions, and also for more general expressions valid for non-Abelian fluxes.

Our calculation is valid for general Calabi-Yau manifolds, includes the effect of worldvolume and bulk fluxes and fully takes into account the non-Abelian nature of the brane fields, thus generalizing and unifying in a single calculation various previous results in the literature. We stress that our calculation also takes into account that the normal bundle of typical divisors is non-trivial (e.g., in the context of the LARGE-volume scenario).
As further novel results, we obtained the 8d potential and the 8d BPS equations in the presence of bulk fluxes. We also discussed in detail how to integrate out KK modes and proposed that controlling related corrections in the 4d effective field theory imposes lower bounds on the Calabi-Yau volume and upper bounds on the brane fields.

This work is intended as a first step in a larger program with the goal of scrutinizing type IIB compactifications with non-commutative D7-brane bound states (T-branes). Such configurations are expected to be a generic feature in type IIB flux compactifications with D7-branes and were moreover proposed in \cite{Cicoli:2015ylx} as an uplift mechanism for de Sitter vacua. It is therefore crucial to understand to what extent T-branes are compatible with the requirements of moduli stabilization and control over dangerous corrections. An extensive analysis of these questions will appear in \cite{Junghans:2026hfm}. A crucial prerequisite for studying moduli stabilization is of course to know the scalar potential. In this paper, we derived a general expression for the D7-brane contribution to this potential from first principles.

A natural way to extend our work would be to push the computation of the brane potential to higher orders in $\alpha'$, i.e., higher orders in the brane fields and/or derivatives (see also \cite{Marchesano:2010bs, Marchesano:2016cqg} for related work). It would also be interesting to explicitly compute KK corrections or 10d backreaction corrections.

Finally, it would be interesting to study the \emph{microscopic} dynamics of T-brane solutions in the presence of bulk fluxes, i.e., not in 4d (where the KK modes are already integrated out) but from the higher-dimensional point of view. Our work  revealed how the 8d potential governing the non-Abelian brane dynamics is modified by bulk fluxes
and can thus serve as a starting point for such analyses.

\section*{Acknowledgments}

This research was funded by the German Research Foundation (Deutsche Forschungsgemeinschaft) under the project number 516370439.

\bibliographystyle{utphys}
\bibliography{groups}

\end{document}